\PassOptionsToPackage{table,xcdraw,prologue}{xcolor}
\documentclass[sigconf,authorversion,nonacm]{acmart}

\usepackage{graphicx}
\usepackage{balance}
\usepackage{bbm}
\usepackage{url}
\usepackage{amsmath}
\usepackage{subcaption}
\usepackage{tcolorbox}
\usepackage[table,xcdraw,prologue]{xcolor}
\usepackage{cleveref}
\usepackage{multirow}
\newcommand{\rev}[1]{{\color{black}#1}}
\newcommand{\revnew}[1]{{\color{black}#1}}

\DeclareMathOperator*{\argmax}{arg\,max}

\usepackage{comment}
\usepackage{enumitem}
\usepackage{xspace}

\newcommand\attrib[1]{{\tt  #1}\xspace}
\newcommand{\at}[1]{\protect\ensuremath{\mathsf{#1}}\xspace}
\newcommand{\gee}{\mathcal{G}}
\newcommand{\eps}{\varepsilon}
\newcommand{\white}{\at{Caucasian}\xspace}
\newcommand{\black}{\at{African American}\xspace}

\newcommand{\systemx}{\at{OmniFair}}
\newcommand{\stitle}[1]{\vspace{1ex}\noindent{\bf #1}}

\usepackage{algorithmicx}
\usepackage{algorithm}
\usepackage{algpseudocode}
\algtext*{EndWhile}
\algtext*{EndIf}
\algtext*{EndFunction}

\algrenewcommand\algorithmicrequire{\textbf{Input:}}
\algrenewcommand\algorithmicensure{\textbf{Output:}}

\newtheorem{definition}{Definition}
\newtheorem{example}{Example}

\newtheorem{lemma}{\rev{Lemma}}

\AtBeginDocument{%
  \providecommand\BibTeX{{%
    \normalfont B\kern-0.5em{\scshape i\kern-0.25em b}\kern-0.8em\TeX}}}

\begin{document}
\fancyhead{}

\title{OmniFair: A Declarative System for Model-Agnostic Group Fairness in Machine Learning}


\author{Hantian Zhang}
\affiliation{\institution{Georgia Institute of Technology}}
\email{hantian.zhang@cc.gatech.edu}

\author{Xu Chu}
\affiliation{\institution{Georgia Institute of Technology}}
\email{xu.chu@cc.gatech.edu}
\author{Abolfazl Asudeh}
\affiliation{\institution{University of Illinois at Chicago}}
\email{asudeh@uic.edu}
\author{Shamkant B. Navathe}
\affiliation{\institution{Georgia Institute of Technology}}
\email{shamkant.navathe@cc.gatech.edu}

\begin{abstract}
Machine learning (ML) is increasingly being used to make decisions in our society.
ML models, however, can be unfair to certain demographic groups (e.g., African Americans or females) according to various fairness metrics. Existing techniques for producing fair ML models either are limited to the type of fairness constraints they can handle (e.g., preprocessing) or require nontrivial modifications to downstream ML training algorithms (e.g., in-processing).

We propose a declarative system \systemx for supporting group fairness in ML. \systemx features a declarative interface for users to specify desired group fairness constraints and supports all commonly used group fairness notions, including statistical parity, equalized odds, and predictive parity. \systemx is also model-agnostic in the sense that it does not require modifications to a chosen ML algorithm. \systemx also supports enforcing multiple user declared fairness constraints simultaneously while most previous techniques cannot. The algorithms in \systemx maximize model accuracy while meeting the specified fairness constraints, and their efficiency is optimized based on the theoretically provable monotonicity property regarding the trade-off between accuracy and fairness that is unique to our system.

We conduct experiments on commonly used datasets that exhibit bias against minority groups in the fairness literature. We show that \systemx is more versatile than existing algorithmic fairness approaches in terms of both supported fairness constraints and downstream ML models. \systemx reduces the accuracy loss by up to $94.8\%$ compared with the second best method. \systemx also achieves similar running time to preprocessing methods, and is up to $270\times$ faster than in-processing methods.


\end{abstract}


\maketitle

\section{Introduction} \label{sec:intro}
\setcounter{figure}{0}

\begin{figure*}[ht]
\centering
\begin{minipage}{0.34\textwidth}
\centering
\includegraphics[width=0.88\columnwidth]{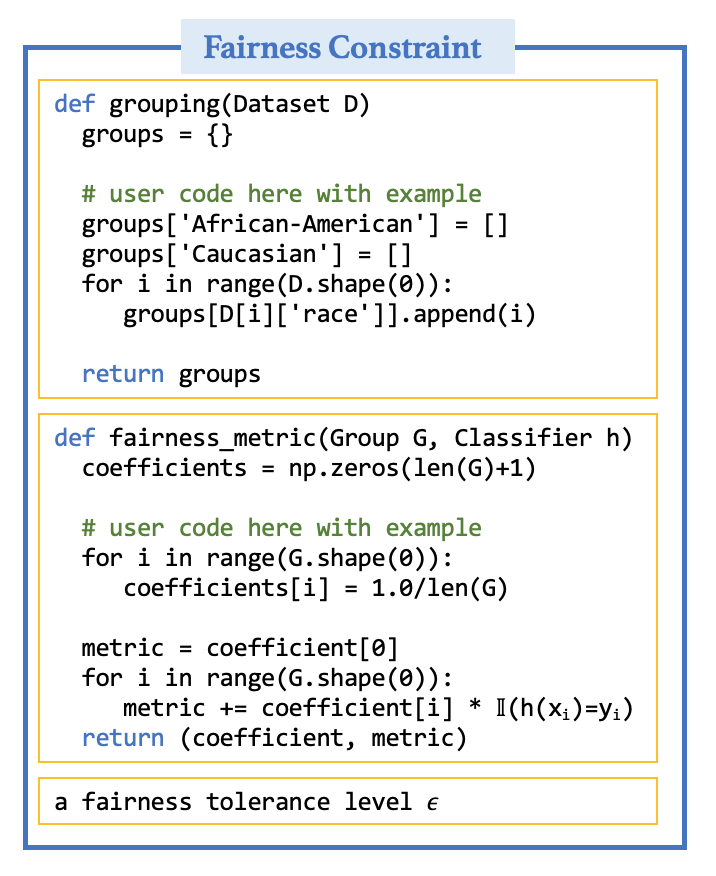}
\vspace{-4mm}
     \caption{The declarative interface.}
    \label{fig:equal_accuracy_specification}
\end{minipage}
\begin{minipage}{0.6\textwidth}
\centering
\captionsetup{type=table} 
\centering
\small
\caption{Comparison with existing algorithmic fairness methods.}
\label{table:baselines}
\vspace{-4mm}
\scalebox{0.85}{
\begin{tabular}{|l|l|l|l|l|l|}
\hline
\multirow{2}{*}{\textbf{Method}}            & \multirow{2}{*}{\textbf{Stage}} & \multirow{2}{*}{\textbf{\begin{tabular}[c]{@{}l@{}}One Fairness \\ Constraint (c.f. \S~\ref{sec:bgnd:GroupFairness})\end{tabular}}} & \multirow{2}{*}{\textbf{\begin{tabular}[c]{@{}l@{}}Multiple\\ Constraints\end{tabular}}} & \multirow{2}{*}{\textbf{\begin{tabular}[c]{@{}l@{}}Constraint\\ Customization\end{tabular}}} & \multirow{2}{*}{\textbf{\begin{tabular}[c]{@{}l@{}}Model \\ Agnostic\end{tabular}}} \\
                                            &                                 &                                                                                                                 &                                                                                          &                                                                                              &                                                                                     \\ \hline
Kamiran et al.~\cite{kamiran2012data}       & Preprocessing                   & SP                                                                                                              & No                                                                                       & No                                                                                           & Yes                                                                                 \\ \hline
Calmon et al.~\cite{calmon2017optimized}    & Preprocessing                   & SP                                                                                                              & No                                                                                       & No                                                                                           & Yes                                                                                 \\ \hline
Zafar et al.~\cite{zafar2017fairness}       & In-Processing                   & \begin{tabular}[c]{@{}l@{}}MR,SP\\ FPR,FNR\end{tabular}                                                            & No                                                                                       & No                                                                                           & No                                                                                  \\ \hline
Celis et al.~\cite{celis2019classification} & In-Processing                   & \begin{tabular}[c]{@{}l@{}}MR,SP\\ FPR,FNR\\ FDR,FOR\end{tabular}                                                  & Yes*                                                                                     & No                                                                                           & No                                                                                  \\ \hline
Agarwal et al.~\cite{agarwal2018reductions} & In-Processing**                   & \begin{tabular}[c]{@{}l@{}}MR,SP\\ FPR,FNR\end{tabular}                                                            & Yes*                                                                                     & No                                                                                           & Yes                                                                                 \\ \hline
\systemx                                    & In-Processing**                   & \begin{tabular}[c]{@{}l@{}}MR,SP\\ FPR,FNR\\ FOR,FDR\end{tabular}                                                  & Yes                                                                                      & Yes                                                                                          & Yes                                                                                 \\ \hline
\end{tabular}
}
\vspace{2mm}
\small{\raggedright Yes* means theoretically yes, but practically difficult to do.}
\vspace{-2mm}

\small{\raggedright In-Processing**  \rev{means that these approaches are considered  in-processing systems as they still need access to the ML algorithms to decide on weights. However, they do not need to modify the ML algorithms, like preprocessing techniques.}}
\end{minipage}
\vspace{-5mm}
\end{figure*}


Machine learning (ML) algorithms, in particular classification algorithms, are increasingly being used to aid decision making in every corner of society. There are growing concerns that these ML algorithms may exhibit various biases against certain groups of individuals. For example, some ML algorithms are shown to have bias against African Americans in predicting recidivism~\cite{flores2016false,dressel2018accuracy}, in  NYPD stop-and-frisk decisions~\cite{goel2016precinct}, and in granting loans~\cite{fernandez2007study}.
Similarly, some are shown to have bias against women in job screening~\cite{dastin2018amazon} and in online advertising~\cite{sweeney2013discrimination}. ML algorithms can be biased primarily because the training data these algorithms rely on may be biased, often due to the way the training data are collected~\cite{olteanu2019social,asudeh2019assessing}.

Due to the severe societal impacts of biased ML algorithms, various research communities are investing significant efforts in the general area of fairness --- two out of five best papers in the premier ML conference ICML 2018 are on algorithmic fairness, the best paper in the premier database conference SIGMOD 2019 is also on fairness~\cite{salimi2019interventional}, and even a new conference ACM FAccT (previously FAT*) dedicated to the topic has been started since 2017.  
One commonly cited reason for such an explosion of efforts is the lack of an agreed mathematical definition of a fair classifier~\cite{fairmlbook,narayanan2018tutorial,verma2018fairness}. As such, many different fairness metrics have been proposed to determine how fair a classifier is with respect to a ``protected group'' of individuals (e.g., \attrib{African-American} or \attrib{female}) compared with other groups (e.g., \attrib{Caucasian} or \attrib{male}), including statistical parity~\cite{dwork2012fairness}, equalized odds~\cite{hardt2016equality}, and predictive parity~\cite{dieterich2016compas}. 

\vspace{-3mm}
\begin{example}\label{runningEx2}


A popular model called COMPAS developed by Northpointe, Inc is used frequently in various stages in the criminal justice system, which predicts a dependent's risk of re-offending. Various studies have attempted to judge how fair the model $h$ is for two groups of individuals, namely, $g_1 =$  \attrib{African Americans} and $g_2 =$ \attrib{Caucasians.} 

The famous Propublica study~\cite{machinebias} concluded that the COMPAS model $h$ is unfair because the probability of an individual being classified as ``high-risk'' depends on their race. Specifically, \attrib{African-American} individuals are more likely to be classified as high-risk, i.e., 
$Pr(h(x) = 1 |g_1) > Pr(h(x) = 1 | g_2)$,
a violation of statistical parity~\cite{dwork2012fairness}. 



Northpointe Inc. later published a response~\cite{machinebias-response} to the Propublica study. They claim that the COMPAS model is indeed fair by demonstrating that the model achieves approximately equal accuracy across different groups, namely, 
$Pr(h(x) = y |  g_1) \simeq Pr(h(x) = y | g_2)$. Similarly, researchers also find that the model approximately equal false omission rate and false discovery rate (also known as predictive parity) across different race groups, namely, $Pr(y=1 \vert g_i, h(x) = 0) \simeq Pr(y = 1 \vert g_j, h(x)=0) $ and 
$Pr(y = 0 \vert g_i, h(x) = 1) \simeq Pr(y = 0 \vert g_j, h(x) = 1) $.


An investigation by US Court~\cite{machinebias-response-court} also deemed the COMPAS model $h$ to be fair because it has both approximately equal false positive rate and equal false negative rate across different groups (also known as equalized odds). In other words, it found that the chance of mistakenly classifying an individual as high-risk does not depend on his/her \attrib{race}, namely, 
$Pr(h(x) = 1 \vert g_i, y=0) \simeq Pr(h(x) = 1 \vert g_j, y=0) $ and 
$Pr(h(x) = 0 \vert g_i, y=1) \simeq Pr(h(x) = 0 \vert g_j, y=1) $.


\end{example}
\vspace{-3mm}
All aforementioned  ``definitions'' of fairness  seem reasonable and what to consider as fair depends on  a particular application scenario and to a beholder, and new definitions keep coming up~\cite{narayanan2018tutorial,verma2018fairness}. 

\stitle{Existing Approaches.} Many techniques have been developed to produce fair models in two categories: \textit{pre-processing} approaches that modify the input training data~\cite{kamiran2012data,zemel2013learning,feldman2015certifying,calmon2017optimized} to remove the correlations between the \textit{sensitive attribute} (e.g., race) and the training  labels $y$ before ML algorithms are applied; and \textit{in-processing} approaches that enforce fairness constraints in the  training process
~\cite{zafar2017fairness,agarwal2018reductions,celis2019classification}.
The main advantage of preprocessing approaches is that they are model-agnostic, and thus users can use ML algorithms as-is; however, they can only handle statistical parity~\cite{zemel2013learning,feldman2015certifying,calmon2017optimized}, 
as other constraints require access to both the prediction $h(x)$ and the ground-truth $y$ at the same time (e.g., equalized odds and predictive parity). In contrast, in-processing techniques can handle a much wider range of fairness constraints and can often achieve a better accuracy-fairness trade-off; however, they usually require non-trivial modifications to the ML training procedures~\cite{zafar2017fairness,celis2019classification}, and hence are usually model-specific (except for~\cite{agarwal2018reductions} as far as we know).


\stitle{Key Components of Our Proposal.} In light of the many current and constantly increasing types of fairness constraints and the drawbacks of existing approaches, we develop \systemx. A comparison between \systemx and existing approaches is shown in Table~\ref{table:baselines}, as we shall elaborate next.

\noindent \emph{(1) Declarative Group Fairness.} Current algorithmic fairness techniques are mostly designed for particular types of group fairness constraint. In particular, preprocessing techniques often only handle statistical parity~\cite{kamiran2012data,calmon2017optimized}. 
While  in-processing techniques generally support more types of constraints, they often require significant changes to the model training process. For example, as shown in Table~\ref{table:baselines}, while Celis et al.~\cite{celis2019classification} supports all constraints shown in \S~\ref{sec:bgnd:GroupFairness}, it is not model-agnostic.
In addition, all current techniques cannot easily be adapted for customized constraints.


\systemx is able to support all the commonly used group fairness constraints. In addition, \systemx features a declarative interface that allows users to supply  future  customized fairness metrics.
As shown in Figure~\ref{fig:equal_accuracy_specification}, a \textit{fairness specification} in \systemx is a triplet $(g, f, \varepsilon)$ with three components: (1) a \at{grouping} function $g$ to specify demographic groups; (2) a \at{fairness\_metric} function $f$ to specify the fairness metric to compare between different groups; and (3) a value $\varepsilon$ to specify the maximum disparity allowance between groups.

Given a dataset $D$, a chosen ML algorithm $\mathcal{A}$ (e.g., logistic regression), and a fairness specification $(g, f, \varepsilon)$, \systemx will return a trained classifier $h$ that maximizes  accuracy on $D$ and, at the same time, ensures that, for any two groups $g_i$ and $g_j$ in $D$ according to the \at{grouping} function $g$, the absolute difference between their fairness metric numbers according to the \at{fairness\_metric} function $f$ is within the disparity allowance $\varepsilon$. 
Figure~\ref{fig:equal_accuracy_specification} shows an example of constraint specification using our interface, which we will explain further in~\S~\ref{sec:declarative}. We will show that our interface can support not only  all common group fairness constraints but also customized ones, \rev{including customized grouping functions such as intersectional groups~\cite{foulds2018intersectional,dwork2018group} and customized fairness metrics.}

\noindent \emph{(2) Example Weighting for Model-Agnostic Property.} The main advantage of preprocessing techniques is that they can be used for any ML algorithm $\mathcal{A}$. The model-agnostic property of preprocessing techniques is only possible when they limit the supported fairness constraints to those that do not involve both the prediction $h(x)$ and the ground-truth label $y$ (i.e., statistical parity). 

Our system \systemx not only supports all constraints current in-processing techniques support,  but also does so in a model-agnostic way. 
Our key innovation to achieve the model-agnostic property is to translate the constrained optimization problem (i.e., maximizing for accuracy subject to fairness constraints) into a weighted unconstrained optimization problem (i.e., maximizing for weighted accuracy). 
Specifically, in order to incorporate the fairness constraints, the original objective function of an ML algorithm $\mathcal{A}$ that maximizes for accuracy is turned into a version that maximizes for weighted accuracy (see \Cref{equ:old_objective} to~\Cref{equ:new_objective} below), where $\lambda$ is a hyperparameter controlling the trade-off between accuracy and disparity of fairness metrics between two groups.
\begin{align} 
 &\textbf{\small{Max for accuracy without fairness constraints:}}  \nonumber \\
  &\max_{\theta} \frac{1}{N} \sum_{i=1}^{N} \mathbbm{1}(h_{\theta}(x_i)=y_i) \label{equ:old_objective} \\
   & \textbf{\small{Max for weighted accuracy with fairness constraints:}}    \nonumber \\
   &\max_{\theta} \frac{1}{N}  \sum_{i=1}^{N} w_i(\lambda, h_{\theta}) \mathbbm{1}(h_{\theta}(x_i)=y_i) \label{equ:new_objective}
\end{align}
This translation is only possible due to the way the fairness metric function is defined, as we shall see later. Solving the weighted objective function, if the example weights $w_i(\lambda, h_{\theta})$ are constant values, does not require any modification to $\mathcal{A}$, since ML algorithms implemented in most ML packages (e.g., scikit-learn) already include an optional parameter to specify training example weights. Even when some ML algorithm implementations do not have the optional parameter, we can simulate weighting by replicating training examples --- for example, a training dataset with two examples with weights 0.4 and 0.6, respectively, can be simulated by replicating the first example two times and the second example three times.

In \S~\ref{sec:single}, we first show in detail how the translation is performed and what the example weights are for the commonly used fairness metrics. 
We then theoretically prove a monotonicity property for both fairness disparity and accuracy with respect to the hyperparameter $\lambda$. 
This unique property allows us to design efficient and effective hyperparameter tuning algorithm to maximize for accuracy while meeting the specified fairness constraint, which involves solving ~\Cref{equ:new_objective} multiple times for different $\lambda$ settings.






\noindent \emph{(3) Supporting Multiple Fairness Constraints.} As discussed before, existing fairness ML techniques already support fewer types of single group fairness constraints than \systemx. 
%
In practice, users may wish to enforce multiple fairness constraints simultaneously. While some in-processing techniques theoretically can support these cases, it is practically extremely difficult to do so, as each fairness constraint is hard-coded as part of the constrained optimization training process (c.f. Table~\ref{table:baselines}).


Our system \systemx can easily support multiple fairness constraints without any additional coding. 
Instead of having one hyperparameter $\lambda$, we will have a vector of fairness hyperparameters $\Lambda$, where each element in $\Lambda$ is used to control one fairness constraint.
Before tuning $\Lambda$ to maximize for accuracy, we need to consider the \textit{feasibility} question under the multi-constraint setting, namely, does there even exist a model that can satisfy all  constraints at the same time. Indeed, prior research has shown theoretically that no model, regardless of which ML algorithms are used, can achieve perfect ($\varepsilon = 0$) statistical parity, perfect equalized odds, and perfect predictive parity for any dataset~\cite{kleinberg2016inherent}.





Given a particular dataset $D$, a particular ML algorithm $\mathcal{A}$, and a particular set of fairness constraints, in \S~\ref{sec:mutiple}, we first prove  that, under some mild conditions, if a given set of fairness constraints can be satisfied simultaneously, there must exist some $\Lambda$ value such that training $\mathcal{A}$ on the weighted dataset produces the feasible model. Since we cannot exhaustively search all $\Lambda$ values (they are real numbers), we devise a practical hill-climbing algorithm for tuning $\Lambda$, which not only is much more efficient than a grid search, but also empirically shows a higher chance of reaching a feasible solution if one exists due to the finer grained search steps we can take.

\stitle{Key features of \systemx}. In summary, we propose \systemx for enforcing group fairness in ML with the following salient features:

\smallskip
\noindent \emph{(1) Versatile.} \systemx is versatile and easy to use --- users only need to specify the desired fairness constraints and the ML algorithm, \systemx is then able to produce a model that maximizes accuracy while meeting the constraints without changing the ML algorithm. \systemx also allows for customized fairness constraints, and can also enforce multiple fairness constraints simultaneously.



\smallskip
\noindent \emph{(2) High Quality.} \systemx consistently outperforms existing methods w.r.t.  accuracy-fairness trade-off. For example, when enforcing a statistical parity constraint with $\varepsilon = 0.03$, \systemx reduces the accuracy loss by up to $94.8\%$ compared with the second best method, \rev{when evaluated on the unseen test set.} \rev{However, we caution that we can only explore the accuracy-fairness trade-off using an input dataset $D$ (which is split into a training and a validation), and the obtained model may not satisfy declared fairness constraints on arbitrary test sets.}



\smallskip
\noindent \emph{(3) Efficient.} \systemx includes efficient algorithms for hyperparameter tuning, which are designed based on theoretical properties unique in our design. We ensure that achieving high quality models of any ML algorithm under declarative group fairness constraints does not come at the expense of high computational overhead. \systemx achieves comparable running time to existing preprocessing methods, and is up to $270\times$ faster than in-processing methods.






\vspace{-4mm}
\section{Related Work} \label{sec:related_work}

\stitle{Systems Support for ML.} With the widespread use of ML analytics, we have seen a surge of systems work from the database community to manage parts of or the whole ML lifecycle with the general goal of democratizing ML~\cite{polyzotis2017data,bailis2017infrastructure,boehm2019data}. Systems such as Snorkel~\cite{ratner2017snorkel} and Goggles~\cite{nilaksh2020goggles} provide declarative programmatic interfaces, in the form of labeling functions and affinity functions, respectively, to allow users to easily express their domain knowledge that is useful for data labeling; systems such as Helix~\cite{xin2018helix} and MLFlow~\cite{zaharia2018accelerating} aim at managing and optimizing the iterative trial-and-error process that is the nature of ML workflow development; and systems such as Vista~\cite{nakandala2020vista} provide declarative feature transfer to allow for jointly analyzing image data and structured data at scale; and systems such as Data Civilizer~\cite{rezig2019data} and ActiveClean~\cite{krishnan2016activeclean} help users deal with data errors in building ML models. \systemx is our systems effort to allow ML users to easily incorporate fairness constraints into ML applications by providing a declarative programming interface while remaining model-agnostic.

\stitle{Algorithmic Fairness Work.}
Table~\ref{table:baselines} shows the major works in algorithmic fairness literature that enforces various group fairness constraints in classification tasks. They can be grouped into two categories: Preprocessing and In-Processing.Preprocessing methods~\cite{kamiran2012data,calmon2017optimized} usually remove the dependency between the sensitive attribute and the target label by transforming the training data via example weighting or feature trasnformation. Preprocessing methods are typically very efficient, as they only need to prepare the training data and do not need to involve the ML model training process, but can only handle statistical parity.


Zafar et al.~\cite{zafar2017fairness}, Agarwal et al.~\cite{agarwal2018reductions}, and Celis et al.~\cite{celis2019classification} are three state-of-the-art in-processing methods, which modify ML algorithms to incorporate fairness constraints. Hence, they usually can handle more constraint types but are usually less efficient, depending on how they solve the constrained optimization problem.
Zafar et al.~\cite{zafar2017fairness} only works for decision boundary based classifiers (e.g., Logistic Regression, SVMs). It uses the covariance of the sensitive attribute and the signed distance between the feature vectors of misclassified data points and the decision boundary to represent fairness constraints and then solve the constrained optimization problem. It does not support predictive parity. 
Celis et al.~\cite{celis2019classification} is the only in-processing work that supports predictive parity. It transforms the optimization problem with predictive parity constraint, which is non-convex, to a specific family of fair classification problem with convex constraints, and hence is not model-agnostic. 

 \rev{
 To the best of our knowledge,
 Agarwal et al.~\cite{agarwal2018reductions} is the only in-processing work that is also model-agnostic. \systemx is different from [4] in three major aspects. First, \systemx supports more constraint types than [4]. In particular, \systemx supports predictive parity and customized constraints whose example weights are parameterized by $h_{\theta}$ while [4] does not. Second, for constraints both \systemx and [4] support, \systemx is much more user-friendly in terms of constraint specification due to our declarative interface. In contrast, users of [4] would need to understand the codebase and implement new constraint classes that connect with the rest of the code, a highly non-trivial task. Third, while both \systemx and [4] leverage the similar idea of Lagrangian multiplier to translate constraint optimization problems to unconstrained optimization problems, our optimization algorithms are much faster because we leveraged the monotonicity property, while [4] used a saddle point finding algorithm, which is known to be difficult to solve.}
 
\rev{Another recent work~\cite{thomas2019preventing} shares the same overall objective as OmniFair, i.e. , making it dramatically easier for users to specify and regulate bias in ML without having to think about how to modify ML training algorithms.  Thomas et al.~\cite{thomas2019preventing} proposes a principled framework for ML researchers or engineers to design new ML algorithms that can take fairness constraints as input. In contrast, \systemx can be used with any existing black-box ML algorithms. }

\vspace{-3mm}
\section{Preliminary} \label{sec:background}

\subsection{Machine Learning Background} \label{sec:mlmodels}
In this paper, we limit our scope to  {\em binary classification}, as almost all current fairness literature does. 
That is, we consider each tuple to include a set of features $x \in \mathbbm{R}^d$ and to be associated with a binary label $y\in \{ 0,1\}$.
Our goal is to learn a classifier $h_\theta(x)$,
where $h:\mathbbm{R}^d\rightarrow \{ 0,1\}$ is the classifier function that is parameterized by  $\theta$.
We assume a dataset $D=\{(x_i,y_i)\}_{i=1}^N$ is used for training the classifier. $h_\theta$ predicts the class label of a query point $x$ as $h_\theta(x)$.

An ML training algorithm $\mathcal{A}$ takes $D$ as input and produces a classifier $h_{\theta}(x)$ by maximizing for empirical accuracy (c.f.~\Cref{equ:old_objective}).
Different ML algorithms optimize for the empirical accuracy differently. Many algorithms use some \textit{loss function} $L(x_i,y_i; \theta)$ as a proxy for the identity function $\mathbbm{1}(h_{\theta}(x_i)=y_i)$, and minimize the empirical loss.
For example, for logistic regression, $L(x_i,y_i; \theta)$ is the logistic loss, while for support vector machines (SVM), it is the hinge loss. On the other hand, some ML algorithms (e.g., decision trees) may not have an explicit loss function, but still optimize for accuracy (e.g., by maximizing for purity of nodes in the tree). 

\vspace{-3mm}
\subsection{Group Fairness Constraints}\label{sec:bgnd:GroupFairness}
While there exist other fairness notions (e.g., individual fairness~\cite{dwork2012fairness} and causal fairness~\cite{russell2017worlds,salimi2019interventional}), group fairness remains the most popular in practice~\cite{machinebias,calmon2017optimized,celis2019classification,feldman2015certifying,hardt2016equality,kamiran2012data}.
An ML model $h(\cdot)$ satisfies some group fairness constraints if the model has equal or similar performance (according to some fairness metrics) on different  groups $\gee$.
We list  major existing group fairness constraints as follows:
\begin{itemize}[leftmargin=*]
    \item {\bf Statistical Parity (SP)} is the most studied group fairness constraint, which makes the independence assumption of the model from demographic groups ($h\,\bot\, \gee$). 
    Under SP~\cite{dwork2012fairness}, the probability of a model outcome is equal or similar across different groups:
    \begin{small}
    \begin{equation}
    \label{equation:dp}
        \forall g_i, g_j \in \gee, Pr(h(x)=1\vert g_i) \simeq	 Pr(h(x)=1\vert g_j)
    \end{equation}
    \end{small}
    \item {\bf False Positive Rate Parity (FPR)} makes the independence assumption of model from demographic groups, conditional to the true labels being 0 ($h\,\bot\, \gee~\vert ~y=0$).
    \begin{small}
    \begin{equation}
    \label{equation:fpr}
         \forall g_i, g_j \in \gee, Pr(h(x) = 1 \vert g_i, y=0) \simeq Pr(h(x) = 1 \vert g_j, y=0) 
    \end{equation}
    \end{small}
    \item  {\bf False Negative Rate Parity (FNR)} is similar to FPR. The only difference is that FNR is conditioned on $y=1$.
    %
    If both FPR and FNR are satisfied, then \textbf{Equalized Odds}~\cite{hardt2016equality} is satisfied.  
    
    \item {\bf False Omission Rate Parity (FOR)} makes the independence assumption of true label from groups, conditional on the negative model prediction ($y\,\bot\, \gee\vert ~h=0$). 
    \begin{small}
    \begin{equation}
    \label{equation:pp_pos}
        \forall g_i, g_j \in \gee, Pr(y = 1 \vert g_i, h(x) = 0) \simeq Pr(y = 1 \vert g_j, h(x) = 0) 
    \end{equation}
    \end{small}
    \item {\bf False Discovery Rate Parity (FDR)} is similar to FOR. The only  difference is that FDR is conditioned on $h(x)=1$.
    If both equal or similar FOR and FDR are satisfied, then \textbf{Predictive Parity}~\cite{dieterich2016compas} is satisfied.  
    
    \item {\bf Misclassification Rate Parity (MR)} equalizes the misclassification rate across different groups. 
    \begin{small}
    \begin{equation}
        \forall g_i, g_j \in \gee, Pr(h(x)=y\vert g_i) \simeq Pr(h(x)=y\vert g_j)
    \end{equation}
    \end{small}
\end{itemize}




\vspace{-3mm}
\section{The Declarative Specification} \label{sec:declarative}

We introduce the declarative interface of \systemx. Our interface not only supports all group fairness constraints presented in \S~\ref{sec:bgnd:GroupFairness}, but also allows users to supply customized constraints.  

\begin{definition} \textbf{Fairness Specification and Fairness Constraint}
\label{def:fairness}
    A \emph{fairness specification} is given by a triplet $(\at{g}, \at{f}, \varepsilon)$. One fairness specification on $D$ induces $|\at{g}(\at{D})|\choose{2}$ \emph{fairness constraints}, each defined over a pair of groups in $\at{g}(\at{D})$. 
    
    A fairness specification is said to be satisfied by a classifier $h$ on $D$ if and only if all induced fairness constraints are satisfied, i.e., 
    $\forall g_i,g_j \in \at{g}(\at{D}),~ |\at{f}(h,g_i) - \at{f}(h,g_j)| \leq \varepsilon$
\end{definition}

\rev{We note that the above definition defines two concepts: fairness specification and fairness constraint. A specification is a triplet $(\at{g}, \at{f}, \varepsilon)$ that users specify. A given specification can induce one or multiple constraints, one for each pair of groups given by $\at{g}$. To specify constraints that require different grouping or metrics, users need to provide multiple specifications. Section~\ref{sec:single} handles a single specification that gives one constraint only, i.e., the grouping function generates only two groups. Section~\ref{sec:mutiple} handles multiple constraints, which may be induced by a single specification that generates multiple groups or multiple specifications. }


\vspace{-1mm}
\stitle{Problem Formulation.} Given a dataset $D$, an ML algorithm $\mathcal{A}$, one or multiple group fairness constraints given by one or multiple fairness specifications, our goal is to obtain an ML model $h_{\theta}$ that maximizes for accuracy, while satisfying given constraint(s).

\rev{\stitle{Discussion:} In practical deployment, we obtain $h_{\theta}$ using an input dataset $D$, which we will split into $D_{train}$ and $D_{val}$ for better generalizability when tuning hyperparameters. We make sure that the model $h_{\theta}$ satisfies the declared constraints on $D_{val}$. However, $h_{\theta}$ may still not satisfy the constraints on arbitrary unseen test sets. }
\vspace{-3mm}
\subsection{Declarative Grouping Function}\label{sec:declarative:group}

Naturally, any declarative group fairness specification must allow users to specify the interested demographic groups.
\begin{definition} (Declarative Grouping Function)\label{def:declarativegroup}
A declarative grouping function \at{g} is a user-defined function that takes a dataset \at{D} as input, and
{\em partitions} the tuples to different groups $\gee$. We implement $\gee$ as a dictionary, in which the keys are group ids and the values are the set of tuples in each group.
\end{definition}



\vspace{-3mm}
\begin{example}
As a toy example, consider a dataset $D=\{t_1,t_2,\cdots t_{10}\}$, where $t_4$, $t_5$, $t_7$, and $t_9$ are 
African American and others are
Caucasian. A user-specified grouping function is 
 shown in Figure~\ref{fig:equal_accuracy_specification}:
The above function partitions $D$ as: $\at{g}(D) = \{\attrib{Caucasian}:[1,2,3,6,8,10], \\ \attrib{African American}:[4,5,7,9]\}$.
\end{example}
\vspace{-2mm}
Note that almost all current group fairness constraints assume that the groups are implicitly induced by a given sensitive attribute (e.g., race). In fact, some preprocessing techniques are applicable only when groups are induced by a sensitive attribute, since they enforce group fairness by removing the correlations between the sensitive attribute and the training data labels~\cite{feldman2015certifying,calmon2017optimized}. 
Our grouping function removes such implicit assumption and allows greater flexibility in declaring groups as we further explain in \S~\ref{sec:3:custom}. 

\vspace{-3mm}
\subsection{Declarative Fairness Metric Function} \label{sec:declarative_metrics}

Designing the interface for specifying fairness metric requires greater consideration. In particular, the fairness metric interface needs to strike a balance between its ability to express various group fairness constraints and the hardness of designing model-agnostic techniques to enforce the declared fairness constraints. 
We propose to express the group fairness metric as a weighted linear summation of the identity function $\mathbbm{1}(h(x_i) = y_i)$. Our intuition is that in order to design a model-agnostic approach for enforcing fairness constraints, we cannot assume any knowledge about how the ML algorithm $\mathcal{A}$ works internally. The only behavior we do know about  $\mathcal{A}$ is that it optimizes for accuracy in the absence of any constraints, and the calculation of accuracy is based on whether each individual prediction is correct, i.e., $\mathbbm{1}(h(x_i) = y_i)$. If we are able to express fairness constraints based on $\mathbbm{1}(h(x_i) = y_i)$, we may be able to optimize for accuracy and fairness simultaneously.



\begin{definition} (Declarative Fairness Metrics)\label{sec:declarative:function}
A fairness metric \at{f} function takes as input a classifier $h$ and a group $g$, and returns $(1+|g|)$ coefficients that specify how the metric is computed: 
\begin{equation}
    \at{f}(h,g) = \sum_{i \in g} c_i\mathbbm{1} (h(x_i)=y_i) + c_0
\end{equation}
\end{definition}


\vspace{-2mm}
Different fairness metrics are specified by different coefficients $c_i$ on each data point in group $g$. 
Our interface supports all existing group fairness constraints (the coefficients for group fairness constraints are summarized in Table~\ref{table:fairnessmetric}), and we can design efficient model-agnostic techniques for enforcing constraints declared this way (c.f. \S~\ref{sec:tech:weigh_translation}). 
We show the derivation of SP as an example, and refer readers to the full report~\cite{Full_Report} for additional examples.
\setlength{\tabcolsep}{1pt}
\begin{table}[th!] 
\centering
\small
\vspace{-3mm}
\caption{\small{Coefficients for different popular group fairness metrics. 
The coefficients $c_i$ for the $i^{th}$ example in a group $g$ are split into two groups based on the label $y_i$.
Note that for the first four metrics, the coefficients are not parameterized by $h(x)$; and for the last two metrics, the coefficients are parameterized by $h(x)$.}}
\label{table:fairnessmetric}
\vspace{-3mm}
\begin{tabular}{|l|l|l|l|}
\hline

& $c_i \vert y_i = 0 $& $c_i \vert y_i = 1 $ & $c_0$ \\ \hline
$\text{MR}$ & $1 / \vert g \vert$ &  $1 / \vert g \vert$ & 0 \\ \hline
$\text{SP}$ & $ - 1 / \vert g \vert$ & $ 1 / \vert g \vert$ & $\vert \{i:i \hspace{-1mm} \in \hspace{-1mm} g,y_i \hspace{-0.7mm} = \hspace{-0.7mm} 0\} \vert / \vert g \vert$ \\ \hline
$\text{FPR}$ & $1 / \vert \{i:i \hspace{-1mm}\in \hspace{-1mm} g, y_i = 0\} \vert$ & 0 & 0 \\ \hline
$\text{FNR}$ & 0 & $1 / \vert \{i:i \hspace{-1mm} \in\hspace{-1mm} g, y_i = 1\} \vert$  & 0 \\ \hline
\hline
$\text{FOR}$ & $1 / \vert \{i: i \hspace{-1mm} \in \hspace{-1mm} g,h(x_i) \hspace{-0.7mm} = \hspace{-0.7mm} 0\} \vert$  & 0 & 0 \\ \hline
$\text{FDR}$  & 0 & $1 / \vert \{i: i \hspace{-1mm} \in\hspace{-1mm} g,h(x_i)=1\} \vert $ & 0 \\
\hline
\end{tabular}
\vspace{-5mm}
\end{table}

\vspace{-3mm}
\begin{example}
\stitle{Expressing SP.}
As shown in ~\Cref{equation:dp}, under SP, the fairness metric is $f(h,g) = Pr(h(x)=1)$, which can be decomposed into two parts, $y_i=0$ and $y_i=1$ as follows:
\begin{equation}
\footnotesize
    \begin{split}
      &f(h,g) = Pr(h(x)=1) \\
      &= Pr(y=1)Pr(h(x)=1|y=1) + Pr(y=0)Pr(h(x)=1|y=0) \\
      &= Pr(y=1)Pr(h(x)=1|y=1) + Pr(y=0)(1-Pr(h(x)=0|y=0)) \\
      &=\frac{1}{\vert g \vert}\sum_{ \{i : i \in g, y_i = 1\}} \mathbbm{1} (h(x_i)=y_i) +\frac{-1}{\vert g \vert}\sum_{ \{i : i \in g, y_i = 0\}} \mathbbm{1} (h(x_i)=y_i) \\
      & \quad  + \frac{\vert \{i : i \in g, y_i = 0\} \vert}{\vert g \vert}
    \end{split}
\end{equation}
Therefore, for data points in $g$ with label $y_i = 1$ (resp. $y_i = 0$), the coefficient is $c_i = \frac{1}{\vert g \vert}$ (resp. $c_i = -\frac{1}{\vert g \vert}$).
We also  have $c_0 = \frac{\vert \{i : i \in g, y_i = 0\} \vert}{\vert g \vert}$.
\end{example}


\vspace{-3mm}
\subsection{Fairness Constraint Customization}\label{sec:3:custom}

In addition to supporting existing group fairness constraints, our declarative interface allows for declaring customized constraints. 


\stitle{Customization of Grouping Function.} The grouping function (\S~\ref{sec:declarative:group}) can easily support intersectional fairness~\cite{foulds2018intersectional,dwork2018group}, where fairness metrics are defined for subgroups defined over the intersection of multiple sensitive attributes, and some richer forms of subgroup fairness~\cite{kearns2019empirical}, where groups are defined based on some classification models. Whatever ways one may desire to form groups can be encoded in the grouping function. 
For example, if both race and gender are considered sensitive attributes in a dataset, then one can write a grouping function that outputs the \attrib{African American female} group, the \attrib{African American male} group, and so on. 


\stitle{Customization of Fairness Metrics.} The abstraction to declarative fairness metric enables a wide range of user-customized metrics defined based on societal norms, as long as the fairness metric can be expressed as a weighted linear combination of the identify function (c.f. Definition~\ref{sec:declarative:function}). 
We provide an example of metric customization.



\begin{example}
\label{ex:cutomized_metric}
A model for binary classification tasks can make two types of errors: false positives and false negatives. In statistics, they are called Type I error and Type II error. In different applications, false positives and false negatives may incur different costs~\cite{berk2018fairness}. 

Users can easily specify a particular cost for the two error types and define a customized fairness metric that calculates the average cost per group. We refer to the full report~\cite{Full_Report} for concrete definition.
\end{example}

\rev{\stitle{Scope and Limitations.} OmniFair currently supports group fairness constraints defined in the context of binary classification problems. In terms of the grouping function, users can write any logic for forming groups as long as the function returns at least two groups of tuples (e.g., intersectional groups defined over multiple sensitive attributes). In addition, the returned groups do not necessarily need to be disjoint. In terms of the fairness metric function, OmniFair can only support those metrics that can be expressed as a linear combination of the identify function (c.f. Definition~\ref{sec:declarative:function}). 

We further note that OmniFair can only constrain the difference of a given metric between a pair of groups to be within a threshold, i.e., $|\at{f}(h,g_i) - \at{f}(h,g_j)| \leq \varepsilon$. For example, we do not support the constraining of the division of a given metric between two groups to be within a threshold, and we do not support constraints expressed between more than two groups.}

\section{Single Fairness Constraint} \label{sec:single}

In this section, we consider a single fairness constraint defined over two groups, and defer the general case of handling multiple constraints to \S~\ref{sec:mutiple}. Specifically, given a dataset $D$, an ML algorithm $\mathcal{A}$, and a fairness specification $(g, f, \varepsilon)$, where applying $g(D)$ yields two groups $g_1$ and $g_2$, we aim to obtain a classifier that maximizes for accuracy while satisfying the fairness constraint.

Without loss of generality, we assume the goal of $\mathcal{A}$ is to learn some model parameters $\theta$, and we use $h_{\theta}$ to denote the classifier. 
To simplify notations, we use $AP(\theta) = \frac{1}{N} \sum_{i=1}^{N} \mathbbm{1} (h_{\theta}(x_i)=y_i)$ to denote the \textit{accuracy part} and $FP(\theta) = \at{f}(h_{\theta},g_1) - \at{f}(h_{\theta},g_2)$ to denote the \textit{fairness part} of the constrained optimization problem. 
We can thus now formally write the constrained optimization problem as:
\begin{equation} \label{equ:original}
\begin{split}
       \max_{\theta} \quad & AP(\theta)  \\
        \text{s.t.} \quad &|FP(\theta)| \leq \varepsilon
    \end{split}
\end{equation}


\vspace{-3mm}
\subsection{Problem Translation} \label{sec:tech:weigh_translation}


\revnew{
Solving the constrained  problem directly in~\Cref{equ:original} would require modifications to $\mathcal{A}$ (as current in-processing techniques do). Instead, we translate ~\Cref{equ:original} into an unconstrained optimization problem by using the \textit{Lagrangian dual function}~\cite{boyd2004convex} $h(\lambda_1,\lambda_2)$ (the last two terms in $h(\lambda_1,\lambda_2)$ come from expanding the $|FP(\theta)| \leq \varepsilon$ constraint in~\Cref{equ:original} into two constraints): 
\begin{equation}\label{equ:h}
\begin{split}
h(\lambda_1,\lambda_2) = &\max_{\theta} AP(\theta) + \lambda_1 (\varepsilon-FP(\theta)) + \lambda_2 ( \varepsilon+FP(\theta)) \\
=& \max_{\theta} AP(\theta) + (\lambda_2 - \lambda_1) FP(\theta) + (\lambda_1 + \lambda_2) \varepsilon
\end{split}
\end{equation}

Following directly from Lagrangian duality (as we will show in the following lemma), $h(\lambda_1,\lambda_2)$ provides an upper bound for~\Cref{equ:original} for any $\lambda_1 >0,  \lambda_2 > 0$ and the bound is proven to be tight under the so-called ``strong duality'' assumption~\cite{boyd2004convex}.


The above direct application of Lagrangian duality would require us to tune two hyperparameters ($\lambda_1$ and $\lambda_2$). We further simplify ~\Cref{equ:h} into ~\Cref{equ:reg2} with only one hyperparameter $\lambda$. We note that this is only possible in our setting because the two constraints are acting on the same term involving $\theta$, i.e., $FP(\theta)$. 
\begin{align} \label{equ:reg2}
        & \max_{\theta} AP(\theta) + \lambda FP(\theta) 
\end{align}
This simplification is justified since we can show that for any $\lambda_1 >0,  \lambda_2 > 0$, there always exists a $\lambda = \lambda_2 - \lambda_1$, such that the optimal solution to  ~\Cref{equ:h} will also be the optimal solution to~\Cref{equ:reg2}.
We summarize the direct application of Lagrangian duality and our simplification as two parts of the following lemma.

\begin{lemma}
\label{theo:feasibility_single}
Assume ~\Cref{equ:original} is feasible and let $\theta^*$ be an optimal solution to~\Cref{equ:original}, then 
\begin{enumerate}
    \item  for any $\lambda_1, \lambda_2>0$, $h(\lambda_1,\lambda_2) \geq AP(\theta^*)$; and under strong duality assumption, $\min_{\lambda_1 > 0,\lambda_2>0} h(\lambda_1,\lambda_2) = AP(\theta^*)$.
    
    \item for any $\lambda_1, \lambda_2>0$, let $\tilde{\theta}$ be an optimal solution to~\Cref{equ:h}, then there exists $\lambda \in \mathbb{R}$ (i.e., $\lambda = \lambda_2 - \lambda_1$) such that $\tilde{\theta}$ also optimizes~\Cref{equ:reg2}; and under strong duality assumption, there exists $\lambda$ such that $\theta^*$ optimizes~\Cref{equ:reg2}.
\end{enumerate}

\end{lemma}


}




\revnew{We refer readers to the full report~\cite{Full_Report} for the  proof. We note that there are multiple different conditions under which strong duality assumption holds (e.g., when the primal problem~\Cref{equ:original} is a linear optimization problem; or when the primal problem is a convex optimization problem and the Slater's condition holds~\cite{boyd2004convex})}. Even without the strong duality assumption, the Lagrangian dual is still commonly used in practice to give a good approximate solution for the primal problem (by optimizing for the upper/lower bound). \revnew{In \systemx's setting, while we cannot solve the primal problem in a model-agnostic way, we are able to solve the dual problem in a model-agnostic way. We also observe that the strong duality assumption has little implication on empirical results --- we used four ML models, three of which have a non-convex loss, i.e., $AP(\theta)$, and \systemx performs well under all four tested models.}

\revnew{Given~\Cref{theo:feasibility_single}, we can essentially solve~\Cref{equ:original} naively in three steps: (i) enumerating all possible $\lambda$ values; (ii) finding the optimal solution to~\Cref{equ:reg2} for every $\lambda$ value; and (iii) out of all the optimal solutions for different $\lambda$ values, pick the one that has the maximal $AP(\theta)$ while $|FP(\theta))| \leq \varepsilon$. In Sec~\ref{sec:tech:tuning}, instead of enumerating all $\lambda$ values, we design a smart algorithm to search for $\lambda$ efficiently.}


\begin{table*}
\caption{Weights for popular group fairness metrics. \rev{Here, we assume $g_1$ and $g_2$ are disjoint for simplicity.}
The weights $w_i$ are split into four groups based on their labels $y_i$ and their group $g$. The weights that do not belong to any of those four groups are 1. In addition, weights may or may not be parameterized by $\theta$.}
\vspace{-4mm}
\label{table:weights}
\centering
\small
\begin{tabular}{|l|l|l|l|l|l|} 
\hline
{\cellcolor[rgb]{0.753,0.753,0.753}}weight    & {\cellcolor[rgb]{0.753,0.753,0.753}}metric & {\cellcolor[rgb]{0.753,0.753,0.753}}\textbf{$w_i \vert y_i = 0, g_1 $}  & {\cellcolor[rgb]{0.753,0.753,0.753}}\textbf{$w_i \vert y_i = 1,g_1 $}  & {\cellcolor[rgb]{0.753,0.753,0.753}}\textbf{$w_i \vert y_i = 0, g_2 $}  & {\cellcolor[rgb]{0.753,0.753,0.753}}\textbf{$w_i \vert y_i = 1,g_2 $}    \\ 
\hline
\multirow{4}{*}{ $w_i(\lambda)$ }          & $\text{MR}$                                         & $1+\lambda N / \vert g_1 \vert$                                   & $1+\lambda N / \vert g_1 \vert$                                  & $1-\lambda N / \vert g_2 \vert$                                   & $1- \lambda N / \vert g_2 \vert$                                            \\ 
\cline{2-6}
                                              & $\text{SP}$                                         & $ 1 - \lambda N / \vert g_1 \vert$                                & $ 1+ \lambda N / \vert g_1 \vert$                                & $1 + \lambda N / \vert g_2 \vert$                                 & $1 - \lambda N / \vert g_2 \vert$                                              \\ 
\cline{2-6}
                                              & $\text{FPR}$                                        & $1-\lambda N / \vert \{i : i \in g_1, y_i = 0\} \vert$                                                                        &    1       &  $1 + \lambda N / \vert \{i : i \in g_2, y_i = 0\} \vert$                                                                       &   1                      \\ 
\cline{2-6}
                                              & $\text{FNR}$                                        &    1     & $1 - \lambda N / \vert \{i : i \in g_1, y_i = 1\} \vert$                                                                       & 1        & $1 + \lambda N / \vert \{i : i \in g_2, y_i = 1\} \vert$                                                                                   \\ 
\hline
\multirow{2}{*}{ $w_i(\lambda, h_{\theta})$ } & $\text{FOR}$                                        & $1 - \lambda N / \vert \{i : i \in g_1, h(x_i) = 0\ \vert$                                                                       &  1     & $1 + \lambda N / \vert \{i : i \in g_2, h(x_i) = 0\} \vert$    & 1                 \\ 
\cline{2-6}
                                              & $\text{FDR}$                                        &  1      & $1 - \lambda N / \vert \{i : i \in g_1, h(x_i) = 1\} \vert$                                                                      & 1       & $1 + \lambda N / \vert \{i : i \in g_2, h(x_i) = 1\} \vert$                                                                                  \\
\hline
\end{tabular}
\vspace{-3mm}
\end{table*}
\vspace{-3mm}
\subsection{Solving~\Cref{equ:reg2} for a Given $\lambda$}
\label{sec:black_box_given_lambda}

\rev{\stitle{Deriving Example Weights.} The key to solving~\Cref{equ:reg2} for any constraints $FP(\theta)$ is to transform it into a weighted regular ML optimization problem. Intuitively, this is possible because both $AP(\theta)$ and $FP(\theta)$ are expressed as linear combinations of $\mathbbm{1} (h_{\theta}(x_i)=y_i)$, and hence the overall $AP(\theta) + \lambda FP(\theta)$ can be converted into a weighted regular ML optimization problem.}

Let us now expand~\Cref{equ:reg2} by plugging in $AP(\theta)$ and $FP(\theta)$. We abbreviate $\mathbbm{1} (h_{\theta}(x_i)=y_i)$ as $\mathbbm{1}_i$, and we use $c_i^{g}$ to denote the coefficient of the $i^{th}$ tuple in group $g$ in the calculation of $f(h,g)$. 
\vspace{-1mm}
\begin{small}
\begin{align} 
\label{equ:reg3}
        & \max_{\theta} AP(\theta) + \lambda FP(\theta) \nonumber \\
        & =  \max_{\theta} \frac{1}{N} \sum_{i=1}^{N} \mathbbm{1}_i  + \lambda \big(\sum_{i \in g_1} (c_i^{g_1} \mathbbm{1}_i + c_0^{g_1}) -  \sum_{i \in g_2} (c_i^{g_2}\mathbbm{1}_i + c_0^{g_2})\big) \nonumber \\
         & = \max_{\theta} \frac{1}{N} \sum_{i=1}^{N} w_i(\lambda, h_{\theta}) \mathbbm{1}(h_{\theta}(x_i) = y_i)
\end{align}
\end{small}
\vspace{-1mm}
The weight $w_i$ for each data point in $D$ can be computed directly from the terms that they are participating in. 
\rev{For the points in $D$ that belong to $g_1$ only, their weights are $w_i = 1 + N \lambda c_i^{g_1}$; and for data points in $g_2$ only, their weights are $w_i = 1 - N \lambda c_i^{g_2}$. For the points that belong to both $g_1$ and $g_2$ (as two groups may be overlapping), their weights are $w_i = 1 + N \lambda c_i^{g_1} - N \lambda c_i^{g_2}$. For points that belong to neither group, their weights are $w_i = 1$.}

We list the weights for  popular group fairness metrics in Table~\ref{table:weights} (by essentially plugging in $c_i$ from Table~\ref{table:fairnessmetric}). As we can see, the example weights derived for different fairness metrics are divided into two categories: those that are not parameterized by $h_\theta(x)$ (e.g., SP) and those that are parameterized by $h_\theta(x)$ (e.g., FOR), depending on whether the coefficients $c_i$ in the fairness metric are parameterized by $h_\theta(x)$ (c.f. Table~\ref{table:fairnessmetric}).

\rev{\stitle{Solving~\Cref{equ:reg3}}.} For the fairness metrics whose induced example weights are not parameterized by $h_{\theta}(x)$, the weights $w_i(\lambda)$ are constant values  given $\lambda$, hence ~\Cref{equ:reg3} can be solved using any black-box ML algorithm $\mathcal{A}$ by providing weighted examples.

For the fairness metrics whose induced example weights are indeed parameterized by $h_{\theta}(x)$, i.e., $w_i(\lambda,h_{\theta})$, we cannot solve ~\Cref{equ:reg3} exactly any more, assuming $\mathcal{A}$ is a black-box. We thus resort to find an approximate solution. Our intuition is that \textit{given $\lambda_1$ and $\lambda_2$ that are extremely close to each other, e.g., $\lambda_2 - \lambda_1 \leq \delta = 0.0001$, the predictions from the optimal models $\theta_1$ (given $\lambda_1$) and $\theta_2$ (given $\lambda_2$) should almost be identical.} Therefore, we can approximate the weights associated with $\lambda_2$ using a close enough model $\theta_1$ derived from a close-enough $\lambda_1$, namely, $w_i(\lambda_2, h_{\theta_2}) \simeq w_i(\lambda_2, h_{\theta_1})$. When $\lambda=0$, we can derive the optimal solution $\theta_0$ by setting all the weights $w_i(0,h_{\theta})$ as 1. 
Therefore, starting from $\lambda = 0$ and taking small incremental steps $\delta$, we can obtain accurate weight estimates, which allows us to solve \Cref{equ:reg2} using  black-box  $\mathcal{A}$. 

The additional complexity for dealing with FOR and FDR (i.e., predictive parity), or in general any group fairness constraint that conditions on the prediction $h_{\theta}(x)$, is not surprising,  because those fairness constraints are non-differentiable.
Indeed, even in the fairness literature of in-processing techniques that modify ML algorithms, supporting predictive parity is only a recent development that offers only approximate solutions~\cite{celis2019classification}. As we shall see, \systemx greatly outperforms ~\cite{celis2019classification} in this case.




As we can see, by rewriting~\Cref{equ:reg2} into~\Cref{equ:reg3}, we can solve it in a model-agnostic way. We emphasize that the above translation is possible due to how we abstracted our fairness metric function $f$, which allows us to merge  the objective of maximizing for accuracy and the objective of satisfying fairness constraints into a weighted unconstrained optimization problem. This was a major motivation for our fairness metric interface in \S~\ref{sec:declarative_metrics}.

\vspace{-4mm}
\subsection{Tuning the Hyperparameter $\lambda$}\label{sec:tech:tuning}

Intuitively, the hyperparameter $\lambda$ in ~\Cref{equ:reg2} controls the trade-off between accuracy and fairness --- each $\lambda$ value will produce an ML model (by solving ~\Cref{equ:reg3}) with a particular accuracy and bias. We thus need to find the optimal $\lambda$ that leads to a model $h_{\theta}$ with the maximum possible $AP(\theta)$ while satisfying $|FP(\theta)| \leq \varepsilon$.


Canonical algorithms for hyperparameter search include grid search and random search that randomly tries some values of $\lambda$. Grid search usually produces models with better performance at the cost of efficiency. In our case, the hyperparameter directly controls the trade-off between accuracy and fairness. In fact, we can theoretically prove a \textit{monotonicity} property for $AP(\theta)$ and $FP(\theta)$ with respect to $\lambda$, which allows us to design efficient hyperparameter tuning procedures that cover the entire space. 
\vspace{-1mm}
\subsubsection{The monotonicity property.} \label{sec:monotonicity_single}We show the {\em monotonicity} of both optimization metrics $AP$ and $FP$ with respect to $\lambda$:
\begin{lemma}
\label{thm:monotonicity_single_lamda} [\textbf{Monotonicity for Single Fairness Constraint}]
Consider two values $\lambda_1$ and $\lambda_2$, where $\lambda_1 < \lambda_2$, and let $\theta_1$ and $\theta_2$ denote two optimal solutions to~\Cref{equ:reg2} given $\lambda_1$ and $\lambda_2$, \revnew{assuming they exist} \rev{when trained on the same training set $D$}, respectively:
\begin{equation}
\small
    \theta_1  = \argmax_{\theta} AP(\theta) + \lambda_1 FP(\theta) \label{equ:lambda_1}
\end{equation}
\begin{equation}
    \small
    \theta_2  = \argmax_{\theta} AP(\theta) + \lambda_2 FP(\theta) \label{equ:lambda_2}
\end{equation}
Then, the following properties hold, where \rev{$AP(\theta_1)$, $AP(\theta_2)$, $FP(\theta_1)$, and $FP(\theta_2)$ are evaluated on the same input training set $D$}:
\begin{equation}
\small
    FP(\theta_1) \leq FP(\theta_2)  \label{equ:fp_monotonic}
\end{equation}
\begin{equation}
\small
    AP(\theta_1) \geq AP(\theta_2)  \text{~~when~~} \lambda_2 > \lambda_1 \geq 0 \label{equ:ap_monotonic_pos}
\end{equation}
\begin{equation}
\small
     AP(\theta_1) \leq AP(\theta_2)  \text{~~when~~} 0 \geq \lambda_2 > \lambda_1 \label{equ:ap_monotonic_neg}
\end{equation}
\end{lemma}




\rev{We defer to the full paper~\cite{Full_Report} for proof. 
We note that the fairness part $FP(\theta)$ is monotonically increasing in the full range of $\lambda$, while the accuracy part $AP(\theta)$ is monotonically increasing in the positive range of $\lambda$, and is monotonically decreasing in the negative range of $\lambda$.} Combining \Cref{equ:ap_monotonic_pos} and~\Cref{equ:ap_monotonic_neg}, we can deduce that the ML model has the maximum accuracy when $\lambda = 0$, which is exactly the original objective function for maximizing accuracy and the model may or may not satisfy the fairness constraint.

\rev{We note that while Lemma~\ref{thm:monotonicity_single_lamda} is only theoretically true on the training set, it nevertheless provides a practically efficient way to search over all possible $\lambda$ values. This is based on the assumption that the training set, the validation set, and the test set are from the same distribution, and hence a model with a higher or lower $AP(\theta)$ or $FP(\theta)$ on the training set would also most likely have a higher or lower $AP(\theta)$ or $FP(\theta)$ on the validation set or the test set.}


\setlength{\textfloatsep}{0pt}

\begin{algorithm}[t]
\caption{Tuning Single $\lambda$}\label{alg:tuning_single_lamda_1}
\begin{small}
\begin{algorithmic}[1]
\Require {Dataset $D$, a fairness constraint $(g, f, \varepsilon)$, an ML Algorithm $\mathcal{A}$}
\Ensure {A fair ML model $h_{\theta}$}

	\State $\theta_0 \gets $ apply $\mathcal{A}$ with $w_i(0)$
	\label{line:alg_lamda_1_start_model_0}
	\State $flag \gets false$ {\bf if}  (weights are parameterized by $\theta$) {\bf else} $true$
	\If{ $|FP(\theta_0)| \leq \varepsilon$}
	    \Return $h_{\theta_0}$ \label{line:alg_lamda_1_end_model_0}
	\EndIf
	\If {$FP(\theta_0) > 0$}
	    \State change the order of $g_1$ and $g_2$ in $FP$
	\EndIf
	    \State $\lambda_{l} \gets 0$ and $\lambda_{u} \gets 1$  \label{line:alg_lamda_1_start_search}
	    \If{$flag = true$}
	        \State $\lambda_{l}, \lambda_{u} \gets $ \Call{ExponentialSearch}{$\lambda_{l}, \lambda_{u}$}
	    \Else
	        \State $\lambda_{l}, \lambda_{u} \gets$  \Call{LinearSearch}{$\lambda_{l}, \delta$}
	        {\tt \small // e.g. $\delta=0.001$}
	    \EndIf
	     \While{$\lambda_{u} - \lambda_{l} \geq \tau$} \label{line:alg_lamda_1_start_binary}
	     {\tt\small // $\tau\rightarrow 0$; e.g. $\tau=0.0001$}
	        \State $\lambda_{m} \gets (\lambda_{u} + \lambda_{u})/2$
	        \If{$flag = true$}
	        \State $\theta_{m} \gets $ apply $\mathcal{A}$ with $w_i(\lambda_m)$ 
	        \Else
	        \State $\theta_{m} \gets $ apply $\mathcal{A}$ with $w_i(\lambda_m, h_{\theta_l})$ 	\EndIf
	        \State {\bf if} $FP(\theta_{m}) < -\varepsilon$ {\bf then} $\lambda_{l} \gets \lambda_{m}$
	       \State {\bf else} $\lambda_{u} \gets \lambda_{m}$
	    \EndWhile 
	    \State \textbf{return} $h_{\theta_{m}}$ \label{line:alg_lamda_1_end_binary}
	
	\State
	
	\Function{ExponentialSearch}{$\lambda_{l},\lambda_{u}$}  \label{line:alg_lamda_1_start_exp}
\State $\theta_{u} \gets $ apply $\mathcal{A}$ with $w_i(\lambda_u)$
\While{$FP(\theta_{u}) < -\varepsilon$}
	        \State $\lambda_{l} \gets \lambda_{u}$
	        \State $\lambda_{u} \gets 2 \times \lambda_{u}$
	        \State $\theta_{u} \gets $ apply $\mathcal{A}$  with $w_i(\lambda_u)$
\EndWhile
\State \textbf{return} $\lambda_{l}, \lambda_{u}$
\EndFunction  \label{line:alg_lamda_1_end_exp}

	\State
	
\Function{LinearSearch}{$\lambda_{l}, \delta$} \label{line:alg_lamda_1_start_lin}
\State $\lambda_{u} \gets \lambda_{l} + \delta$
\State $\theta_{u} \gets $ apply $\mathcal{A}$ given $ \lambda_{u}$
\While{$FP(\theta_{u}) < -\varepsilon$}
	        \State $\lambda_{l} \gets \lambda_{u}$
	        \State $\theta_{l} \gets \theta_{u}$
	        \State $\lambda_{u} \gets \lambda_{l} + \delta$
 	        \State $\theta_{u} \gets $ apply $\mathcal{A}$  with $w_i(\lambda_u, \theta_l)$
\EndWhile
\State \textbf{return} $\lambda_{l}, \lambda_{u}$
\EndFunction \label{line:alg_lamda_1_end_lin}

\end{algorithmic}
\end{small}
\end{algorithm}

\subsubsection{Algorithm for Tuning $\lambda$} \label{sec:tuning_lambda_validation}

\Cref{alg:tuning_single_lamda_1} describes the procedure for tuning $\lambda$, which consists of three stages. 


\medskip
\noindent \emph{(1) Training a Model with $\lambda = 0$ (Lines~\ref{line:alg_lamda_1_start_model_0} to~\ref{line:alg_lamda_1_end_model_0}).}  We first train an ML model $\theta_0$ without any fairness constraint, i.e., $\lambda = 0$. If $\theta_0$ already satisfies the fairness constraint, i.e., $|FP(\theta_0)| \leq \varepsilon$, we can simply return $h_{\theta_0}$, which has the maximum $AP(\theta_0)$ according to~\Cref{thm:monotonicity_single_lamda}. 
If $FP(\theta_0) > 0$, we switch the order between the two groups $g_1$ and $g_2$ to make sure $FP(\theta_0) < - \varepsilon$.
This means that we need to search the positive values for $\lambda$, which we describe next.

\medskip
\vspace{-1mm}
\noindent \emph{(2) Bounding $\lambda$ (Lines~\ref{line:alg_lamda_1_start_exp} to~\ref{line:alg_lamda_1_end_exp}).} 
While we know the optimal $\lambda$ must be a positive value, i.e., $\lambda_{lower} = 0$, we do not know its upper bound. 

\emph{(2.1) Exponential Search (Lines~\ref{line:alg_lamda_1_start_exp} to~\ref{line:alg_lamda_1_end_exp}).}  When the weights are not parameterized by $\theta$, then $w_i(\lambda)$ will be constant values for a given $\lambda$ value. We thus employ an exponential search procedure to quickly locate the upper bound $\lambda_{upper}$ by initially setting $\lambda_{upper} = 1$ and gradually doubling it until  $FP(\theta_{u}) \geq -\varepsilon$. 

\emph{(2.2) Linear Search (Lines~\ref{line:alg_lamda_1_start_lin} to~\ref{line:alg_lamda_1_end_lin}).}  When the example weights are parameterized by $\theta$ (e.g., FOR and FDR), the example weights are no longer constant values for a given $\lambda$, and hence we cannot use exponential search to quickly identify the bounds for $\lambda$. As discussed in \S~\ref{sec:black_box_given_lambda}, we have to resort to a linear search procedure for obtaining the approximate weights. Specifically, we start from $\lambda = 0$ and as we take small incremental steps $\delta$, we can obtain accurate weight estimates, which will allow us to solve \Cref{equ:reg3} using the black-box ML algorithm $\mathcal{A}$.

\noindent \emph{(3) Binary Search for the Optimal $\lambda$ (Lines~\ref{line:alg_lamda_1_start_binary} to~\ref{line:alg_lamda_1_end_binary}).} Because of  the  monotonicity property of $FP(\theta)$ in \Cref{equ:fp_monotonic}, we know that there must exist a $\lambda$ in $[\lambda_{lower}, \lambda_{upper}]$ that satisfies the fairness constraint. We can thus do a binary search to reach the smallest $\lambda$ that satisfies the fairness constraint, which must  have the maximum accuracy due to the property of $AP(\theta)$ in ~\Cref{equ:ap_monotonic_pos}.


\rev{\stitle{Use of Validation Set for Generalizability.} For a given ${\mathcal{A}}$, we treat $\lambda$ as a hyperparameter that controls accuracy-fairness trade-off. To avoid overfitting $\lambda$ on the training set, we 
follow the standard practice in tuning hyperparameters in ML and 
use a separate validation set to evaluate the goodness of a $\lambda$ value. In particular, we randomly split $D$ into a $D_{train}$ and a small $D_{val}$. We use $D_{train}$ for training ML models for different $\lambda$ values (i.e., when applying $\mathcal{A}$), and use $D_{val}$ for evaluating ML models (i.e., when calculating $FP(\theta)$ and $AP(\theta)$). Our goal is to pick a value for $\lambda$ such that the model trained with that $\lambda$ value would have the maximum validation accuracy while satisfying the constraint on the validation set. }

\vspace{-2mm}
\rev{While the use of a separate validation set when tuning $\lambda$ helps with the generalizability of the trained model, there is still no guarantee that the trained model will satisfy the declared fairness constraints on arbitrary unseen test sets. However, a larger validation set is better at ensuring that the accuracy and fairness numbers obtained on the validation set are close to those evaluated on a unseen test set, as shown empirically in Figure~\ref{fig:validation}.}

\vspace{-3mm}
\section{Multiple Fairness Constraints} \label{sec:mutiple}

\revnew{
We consider the multiple fairness constraints case in this section. Given $k$ constraints, the constrained optimization problem becomes: 
\vspace{-3mm}
\begin{small}
\begin{equation} \label{equ:original_multi}
\begin{split}
      \max_{\theta} \quad & AP(\theta)  \\
        \text{s.t.} \quad &|FP_i(\theta)| \leq \varepsilon \quad \forall i \in \{1..k\}
    \end{split}
\end{equation}
\end{small}
The Lagrangian dual function~\cite{boyd2004convex} of~\Cref{equ:original_multi} is:
\vspace{-2mm}
\begin{small}
\begin{equation}\label{equ:h_multi}
\begin{split}
h(\Lambda_1,\Lambda_2) = &\max_{\theta} AP(\theta) + \sum_{i=1}^{k} \lambda_{1i} (\varepsilon-FP_i(\theta)) + \sum_{i=1}^{k} \lambda_{2i} ( \varepsilon+FP_i(\theta)) \\
=& \max_{\theta} AP(\theta) + \sum_{i=1}^{k} (\lambda_{1i} - \lambda_{2i}) FP(\theta) + \sum_{i=1}^{k}(\lambda_{1i} + \lambda_{2i}) \varepsilon
\end{split}
\end{equation}
\end{small}
,where $\Lambda_1=\langle\lambda_{11},\lambda_{12}...,\lambda_{1k}\rangle$ and $\Lambda_2=\langle\lambda_{21},\lambda_{22}...,\lambda_{2k}\rangle$.

$h(\Lambda_1,\Lambda_2)$ provides an upper bound for~\Cref{equ:original_multi} for any $\Lambda_1 >0,  \Lambda_2 > 0$ and the bound is tight under strong duality~\cite{boyd2004convex}. We can further simplify ~\Cref{equ:h_multi} into ~\Cref{equation:fairness_opt_multiple} by using half of the hyperparameters, namely, $\Lambda=\langle\lambda_1,\lambda_2...,\lambda_k\rangle$, similar to the single-constraint setting.
}

\vspace{-3mm}
\begin{small}
\begin{align}
&  \max_{\theta} AP(\theta) +  \sum_{i=1}^{k} \lambda_i\, FP_i(\theta)  \label{equation:fairness_opt_multiple} \\
& = \max_{\theta} \frac{1}{N} \sum_{i=1}^{N} w_i(\Lambda, h_{\theta}) \mathbbm{1}(h_{\theta}(x_i) = y_i) \label{equation:fairness_opt_multiple_weighted} 
\end{align}
\end{small}
The transformation from ~\Cref{equation:fairness_opt_multiple} to  ~\Cref{equation:fairness_opt_multiple_weighted}
is similar to what we did in \S~\ref{sec:single}, which allows us to use any black-box ML algorithm $\mathcal{A}$ by adjusting example weights. The  difference is that the weight for every example is now parameterized by $\Lambda$ and $\theta$.

\stitle{Discussion on Feasibility.} Given multiple fairness constraints, there may not exist an ML model that satisfies all constraints. Indeed, a well-known prior research has shown theoretically that no model, regardless of which ML algorithms are used, can achieve perfect ($\varepsilon = 0$) SP, FNR, and FOR, for any dataset~\cite{friedler2016possibility}. 
\revnew{
Fortunately, we can show that our approximation of using~\Cref{equation:fairness_opt_multiple} still works with multiple constraints, if the origianl problem is feasible.

\begin{lemma}
\label{th:impossibility}
Assume ~\Cref{equation:fairness_opt_multiple} is feasible, let $\theta^*$ be an optimal solution to ~\Cref{equation:fairness_opt_multiple}, then
\begin{enumerate}
    \item for any $\Lambda_1, \Lambda_2>0$, $h(\Lambda_1,\Lambda_2) \geq AP(\theta^*)$; and under strong duality assumption $\min_{\Lambda_1 > 0,\Lambda_2>0} h(\Lambda_1,\Lambda_2) = AP(\theta^*)$.
    \item for any $\Lambda_1, \Lambda_2>0$, let $\tilde{\theta}$ be an optimal solution to~\Cref{equ:h_multi},  then there exists $\Lambda \in \mathbb{R}^k$ (i.e., $\Lambda = \Lambda_2 - \Lambda_1$) such that $\tilde{\theta}$ also optimizes~\Cref{equation:fairness_opt_multiple}; and under strong duality assumption, there exists $\Lambda$ such that $\theta^*$ optimizes~\Cref{equation:fairness_opt_multiple}.
\end{enumerate}
\end{lemma}
}


\revnew{~\Cref{th:impossibility} generalizes ~\Cref{theo:feasibility_single} from single constraint to multiple constraints, and the proof is in the full report~\cite{Full_Report}.}
Therefore, similar to the case of tuning $\lambda$ for solving \Cref{equ:original}, we need to tune $\Lambda$ in~\Cref{equation:fairness_opt_multiple_weighted} for solving~\Cref{equ:original_multi}.

\vspace{-2mm}
\subsection{Monotonicity for Multiple Constraints}

For a single constraint, we used a monotonicity property in \Cref{thm:monotonicity_single_lamda}. We now introduce a similar property for multiple constraints. 

\begin{lemma}
\label{thm:monotonicity_multiple_lamda} [\textbf{Marginal Monotonicity for Multiple Fairness Constraints}]
Consider two  settings $\Lambda_1$ and $\Lambda_2$ that differ only in the $j^{th}$ dimension, namely, $\Lambda_1[j]  < \Lambda_2[j]$ and $\Lambda_1[i]  = \Lambda_2[i]$ for all $i \neq j$. 
Let $\theta_1$ and $\theta_2$ denote the optimal solution to ~\Cref{equation:fairness_opt_multiple}  given $\Lambda_1$ and $\Lambda_2$, \rev{when trained on the same training set $D$} respectively, namely, 
\vspace{-3mm}
\begin{equation}
\small
    \theta_1 = \argmax_{\theta} AP(\theta) + \sum_{i=1}^{k} \Lambda_1[i] FP_i(\theta) \label{equ:lambda_multi_1}
\end{equation}
\vspace{-3mm}
\begin{equation}
\small
     \theta_2 = \argmax_{\theta} AP(\theta) + \sum_{i=1}^{k} \Lambda_2[i] FP_i(\theta) \label{equ:lambda_multi_2}
\end{equation}
Then, we have the following monotonicity property, \rev{where $FP_j(\theta_1)$ and $FP_j(\theta_2)$ are evaluated on the same training set $D$,}
\begin{equation}
\small
    FP_j(\theta_1) \leq FP_j(\theta_2)  \label{equ:fp_monotonic_multiple_lambda}
\end{equation} 
\end{lemma}

\rev{The proof of \Cref{thm:monotonicity_multiple_lamda} is similar to that of \Cref{thm:monotonicity_single_lamda} by treating $AP(\theta) + \sum_{i=1, i \neq j}^{k} \Lambda_1[i] FP_i(\theta)$ and $FP_j(\theta)$ in \Cref{thm:monotonicity_multiple_lamda} as the old $AP(\theta)$ and $FP(\theta)$ in \Cref{thm:monotonicity_single_lamda}, respectively.}


Note that we have a ``weaker'' version of the monotonicity property in this case. 
We can only show that the $FP_j(\theta)$ is monotonically increasing with respect to $j^{th}$ dimension in $\Lambda$ given every other dimension stays fixed, and we can provide no guarantees for the $AP(\theta)$ or other $FP_i(\theta), i \neq j$.  

\stitle{Satisfactory regions.} Consider a hyperparameter setting $\Lambda$ and the $j^{th}$ fairness constraint to be  $|FP_j(\theta)| = 0$, the derived solution $\theta$ (by solving~\Cref{equation:fairness_opt_multiple_weighted}) may or may not satisfy $|FP_j(\theta)| = 0$. Suppose $|FP_j(\theta)| = 0$ is not satisfied, according to the marginal monotonicity property, we can always fix $\lambda^-_{j}$ (which denotes all but the $j^{th}$ dimension of $\Lambda$) and update $\lambda_j$ such that $|FP_j(\theta)| = 0$. In other words, for any setting of $\lambda^-_j$, we can always find a corresponding setting for $\lambda_j$ such that $FP_j(\theta) = 0$. We call all such $\Lambda$ settings the \textit{zero-satisfactory region} for $|FP_j(\theta)| = 0$. 

Now let us consider the general $j^{th}$ constraint being $|FP_j(\theta)| \leq \varepsilon$. Similarly, for any setting of $\lambda^-_j$, we can find \textit{a continuous range of possible settings} for $\lambda_j$, such that $|FP_j(\theta)| \leq \varepsilon$. We call all such $\Lambda$ settings the \textit{satisfactory region} for $|FP_j(\theta)| \leq \varepsilon$.

Geometrically, one can observe from marginal monotonicity property that every zero-satisfactory region for $FP_j(\theta) = 0$ is a $(k-1)$ dimensional curved hyperplane that intersects each axis-parallel line in exactly one point, and every satisfactory region for $FP_j(\theta) \leq \varepsilon$ is the intersection of two ``parallel'' half-spaces each identified by a $(k-1)$ dimensional curved hyperplane.

\begin{figure}[t!]
        \centering
        \includegraphics[scale=0.18]{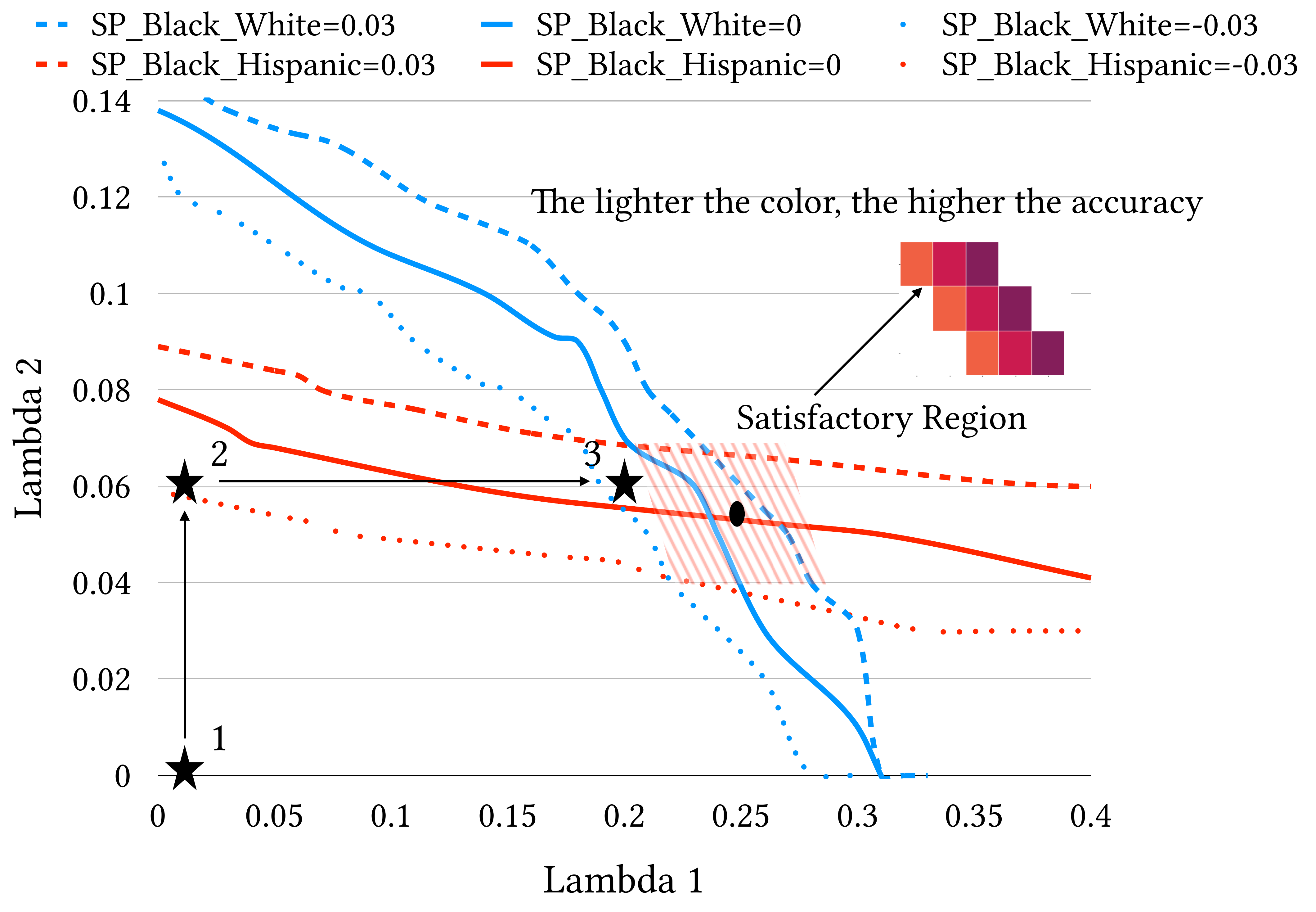}
        \vspace{-5mm}\caption{\footnotesize Zero-satisfactory lines for SP over three demographic groups \white, \black, and \at{Hispanic}.}
        \label{fig:Zline1}
\end{figure}

\vspace{-2mm}
\begin{example}
Figures~\ref{fig:Zline1} shows the satisfactory regions for an experiment on COMPAS dataset, with two fairness constraints ($k=2$). Both constraints enforce statistical parity, one between African American and Caucasian groups and the other between African American and Hispanic groups; both have a disparity allowance of $\varepsilon = 0.03$. 

The solid blue line depicts the zero-satisfactory region for the first constraint, and the region between two blue dashed lines depict the satisfactory region for it. Similarly, the zero-satisfactory region and the satisfactory region for the second constraint is shown in orange. 

All $\Lambda$ settings in the intersection between the blue band the the red band are are thus feasible solutions that satisfy both constraints. 





\vspace{-3mm}

\end{example}

\vspace{-2mm}

\subsection{Algorithm for Tuning  $\Lambda$}
The (hypothetical) baseline solution for finding proper $\Lambda$ is to search the entire space of possible values for $\Lambda$. We can do a grid search to do that. However, doing a grid search could be extremely slow especially when there are many constraints.


\begin{algorithm}[ht]
\caption{Hill-Climbing}\label{alg:greedy}
\small
\begin{algorithmic}[1]
\Require {Dataset $D$, a set of $k$ fairness constraint, and an ML Algorithm $\mathcal{A}$}
\Ensure {A fair ML model $h_{\theta}$}

	\State $\Lambda\gets[0,\cdots,0]$ 
	\State $\theta \gets $ apply $\mathcal{A}$ with $w_i(0, -)$
    \While{$\exists~ i,  \text{  s.t.  } \vert FP_i(\theta) \vert > \varepsilon_i$}
    \State $i \gets \argmax_{k} \vert \vert FP_k(\theta) \vert - \varepsilon_k \vert $
            \State $\theta \gets$ call Algorithm~\ref{alg:tuning_single_lamda_1} to tune for the i-th fairness constraint, while fixing $\Lambda[j], \forall j\neq i$ to their current values
    \EndWhile
    \State {\bf return} $\Lambda$ {\bf if} ($\Lambda\in$ intersection of all satisfactory regions) {\bf else} "Not found after $5k$ iterations"
\end{algorithmic}
\end{algorithm}

Following the marginal monotonicity property and our observations about the shape of satisfactory region, we design a marginal hill-climbing algorithm for parameter tuning when there are multiple fairness constraints.
At a high-level, the marginal hill-climbing algorithm (Algorithm~\ref{alg:greedy}) uses the satisfactory regions as the guidance to move toward their intersection.
That is, at every step, instead of tuning all dimensions, it picks 
a marginal $\lambda^-_j$ and tunes $\lambda_j$ such that the $j^{th}$ constraint is satisfied. While one can naively iterate over all dimensions in a round-robin fashion, at every step we pick the dimension whose fairness constraint is violated the most for faster convergence (Line 4). For a marginal $\lambda^-_j$ at every iteration, the algorithm invokes Algorithm~\ref{alg:tuning_single_lamda_1} for tuning $\lambda_j$ so that the $j^{th}$ constraint is satisfied. This process continues until all constraints are satisfied or a predefined number of iterations are reached (we use 5k in our experiments, where $k$ is the number of constraints).   

Note that the above procedure is implicitly maximizing the impact of loss of model accuracy. This is because, at every iteration,  the algorithm satisfies the $j^{th}$ constraint to its minimum (by invoking Algorithm~\ref{alg:tuning_single_lamda_1}). We empirically do observe that the smaller the update we do on $\lambda_j$, the less impact it has on the accuracy. 





\vspace{-2mm}
\begin{example}
Figure~\ref{fig:Zline1} shows an instance of running Algorithm~\ref{alg:greedy}. Initially, we have $\Lambda_1 = [0,0]$ (the star labeled as 1). The algorithm picks the second dimension (fixing the first dimension) and arrives at $\Lambda_2 = [0, 0.06]$ (the  star labeled as 2). It next picks the first dimension (fixing the second dimension) and arrives at $\Lambda_3 = [0.18,0.06]$ (the star labeled as 3), which satisfies both constraints.
Notice that for all feasible solutions, i.e., in the intersection of both satisfactory regions, the closer $\Lambda$ is to the origin point, the higher the model accuracy. 

\end{example}

\vspace{-2mm}
While tuning the $j^{th}$ dimension can certainly satisfy the $j^{th}$ constraint, it may also impact other dimensions, in particular, it may cause other constraints that had previously been satisfied to be violated again. 
Indeed, we cannot prove that this greedy hill-climbing algorithm will always converge, and hence we set a maximum iteration threshold ($5k$). However, the experiments in \S~\ref{sec:exp_multi} show that empirically, Algorithm~\ref{alg:greedy} can actually recover feasible solutions with a higher probability than grid search with a reasonable step size. We attribute this surprisingly good performance to two factors: (1) at every iteration, though we cannot guarantee tuning $j^{th}$ constraint will not impact other constraints, we indeed are minimizing the potential impact by satisfying the $j^{th}$ constraint to the minimum degree; and (2) Algorithm~\ref{alg:greedy} invokes Algorithm~\ref{alg:tuning_single_lamda_1} for tuning $j^{th}$ dimension, which uses a binary search procedure that can search for extremely fine-grained values ($\tau=0.0001$ in Algorithm~\ref{alg:tuning_single_lamda_1}).

\vspace{-2mm}
\section{Experiments} \label{sec:exp}
We run  experiments to evaluate the effectiveness and efficiency of OmniFair in comparison to existing approaches. 

\vspace{-4mm}
\setlength{\tabcolsep}{2pt}
\begin{table}[h]
    \centering
    \small
    \caption{\small Datasets used}
    \label{tab:datasets}
    \vspace{-3mm}
    \begin{tabular}{|l|r|r|l|l|}
    \hline
        Dataset & \# Rows & \# Attrs & Sens. Attr & Task \\ \hline
        Adult~\cite{Dua:2019} & 48842 & 18 & sex & To predict if Income > 50k \\ \hline 
        Compas~\cite{machinebias} & 11001 & 10 & race& To predict recidivism \\ \hline
        LSAC~\cite{wightman1998lsac} & 27477 & 12 & race& To predict if bar exam is passed \\ \hline
        Bank~\cite{moro2014data} & 30488 & 20 & age& To predict if marketing works\\ \hline
    \end{tabular}
\end{table}
\vspace{-3mm}
\setlength{\tabcolsep}{4pt}
\begin{table*}[]
\caption{Accuracy drop compared with no fairness constraints when $\eps=0.03$ under SP. NA(1) means that no hyperparameter setting can be found to satisfy the SP constraint when $\eps=0.03$. NA(2) means that the classifier is not supported. NA(2)* means that theorectically it could be supported, but CMA-ES does not provide a clean API for invoking the ML algorithm. }
\label{table:accuracy}
\vspace{-3mm}
\scalebox{0.7}{
\begin{tabular}{|l|l|l|l|l|l|l|l|l|l|l|l|l|l|l|l|l|l|l|l|l|}
\hline
&  \multicolumn{5}{c|}{COMPAS} & \multicolumn{5}{c|}{Adult} & \multicolumn{5}{c|}{LSAC}& \multicolumn{5}{c|}{\rev{Bank}}\\ \hline
Algorithm & LR & RF& XGB  &\rev{NN}&\rev{CMA-ES}& LR & RF& XGB&\rev{NN}&\rev{CMA-ES} & LR & RF& XGB &\rev{NN}&\rev{CMA-ES} & LR & RF& XGB &\rev{NN}&\rev{CMA-ES}  \\ \hline
\systemx &-1.2\%& -0.8\%& -0.7\% &-1.2\%&NA(2)* &\textcolor{red}{-2.1\%} &\textcolor{red}{-1.9\%}&\textcolor{red}{-1.7\%}&\textcolor{red}{-1.7\%}&NA(2)*&-0.3\%&\textcolor{red}{-0.3\%}&-0.4\% &\textcolor{red}{-0.1\%}&NA(2)*&-0.1\% &\textcolor{red}{-0.3\%} & -0.2\%&\textcolor{red}{-0.1\%}&NA(2)*\\ \hline
 Kamiran et al.~\cite{kamiran2012data} & -2.5\% & -1.3\%& -1.2\%& -1.5\%&NA(2)*&-2.7\%  & -2.3\%&-1.8\%&-1.9\%&NA(2)*&-0.4\%&-5.6\%&-2.2\% &-0.4\%&NA(2)*&\textcolor{red}{+0.1\%}&-1.2\%&-0.2\%&-0.3\%&NA(2)*\\ \hline
 Calmon et al.~\cite{calmon2017optimized}& -1.8\% & \textcolor{red}{-0.5\%} & \textcolor{red}{-0.3\%} &\textcolor{red}{-0.9\%}&NA(2)*& -3.7\% & -3.1\% & -3.0\%&-2.4\%&NA(2)&NA(1)&NA(1)&NA(1)&NA(1)&NA(2)*&NA(1)&NA(1)&NA(1)&NA(1)&NA(2)* \\ \hline
 Zafar et al.~\cite{zafar2017fairness}& \textcolor{red}{-0.9\%}& NA(2)&NA(2)&NA(2) &NA(2) &-2.2\%& NA(2)&NA(2) &NA(2)&NA(2)& \textcolor{red}{-0.2\%}& NA(2)&NA(2) & NA(2)&NA(2)&-0.1\%&NA(2) & NA(2)&NA(2)&NA(2) \\ \hline
 Celis et al.~\cite{celis2019classification} &NA(1) & NA(2)& NA(2) &NA(2)& NA(2)&NA(1)& NA(2)& NA(2) & NA(2)&NA(2)&NA(1)&NA(2) &NA(2)&NA(2) &NA(2)&NA(1) &NA(2)&NA(2) &NA(2) &NA(2)\\ \hline
 Agarwal et at.~\cite{agarwal2018reductions}& -2.4\%& -1.2\% & -2.0\% &-1.8\%&NA(2)*& -2.8\% & -2.2\%&-2.0\%&-2.0\%&NA(2)*&-0.6\%&-5.8\%&\textcolor{red}{-0.2\%} &-0.5\%&NA(2)*&-0.1\% &\textcolor{red}{-0.3\%}&\textcolor{red}{-0.0\%}& \textcolor{red}{-0.1\%}& NA(2)*\\  \hline
 \rev{Thomas et at.~\cite{thomas2019preventing}}&NA(2)& NA(2) & NA(2) &NA(2)& \textcolor{black}{-1.1\%} &NA(2)& NA(2)&NA(2)&NA(2)&\textcolor{black}{-1.7\%} &NA(2)&NA(2)&NA(2)&NA(2)&\textcolor{black}{-0.4\%} & NA(2) &NA(2)&NA(2)&NA(2)&\textcolor{black}{-0.1\%} \\\hline
\end{tabular}
}\end{table*}
\setlength{\tabcolsep}{2pt}

\vspace{-3mm}
\begin{figure*}[ht]
\noindent\begin{minipage}{0.29\textwidth}
\centering
\vspace{-3mm}\caption{\small Accuracy and  bias on the test set of the COMPAS dataset when varying the validation set size.} 
\label{fig:validation}
\includegraphics[width=\linewidth]{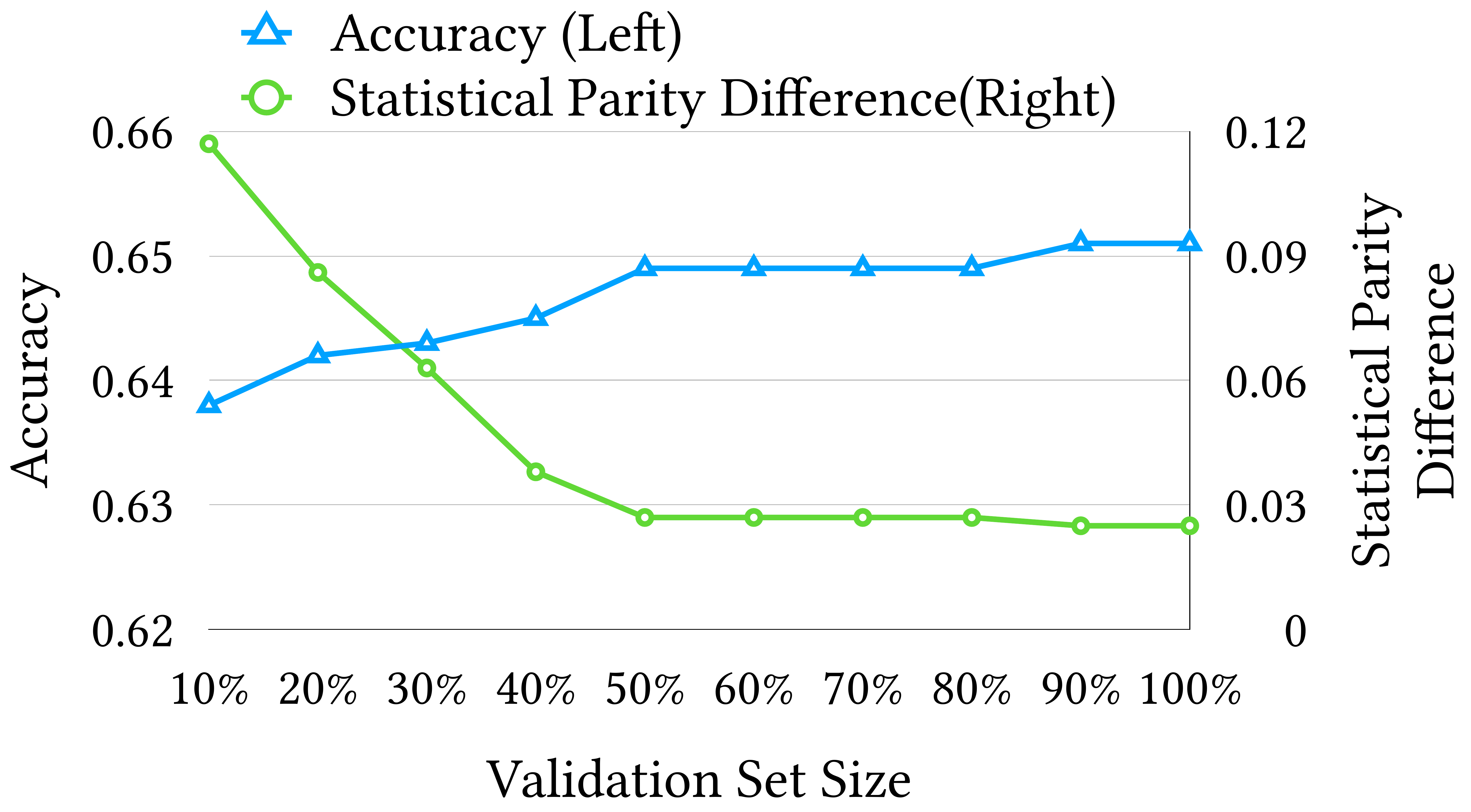}
\vspace{-5mm}
\end{minipage}
\hfill
\noindent\begin{minipage}{0.70\textwidth}
\centering
\vspace{-3mm}\caption{\small The trade-off between fairness metric and accuracy with (a)LR, (b)RF, and \rev{(c)LR (ROC AUC)} on Adult dataset.} 
\label{fig:single_fairness_adult}
\includegraphics[width=\linewidth]{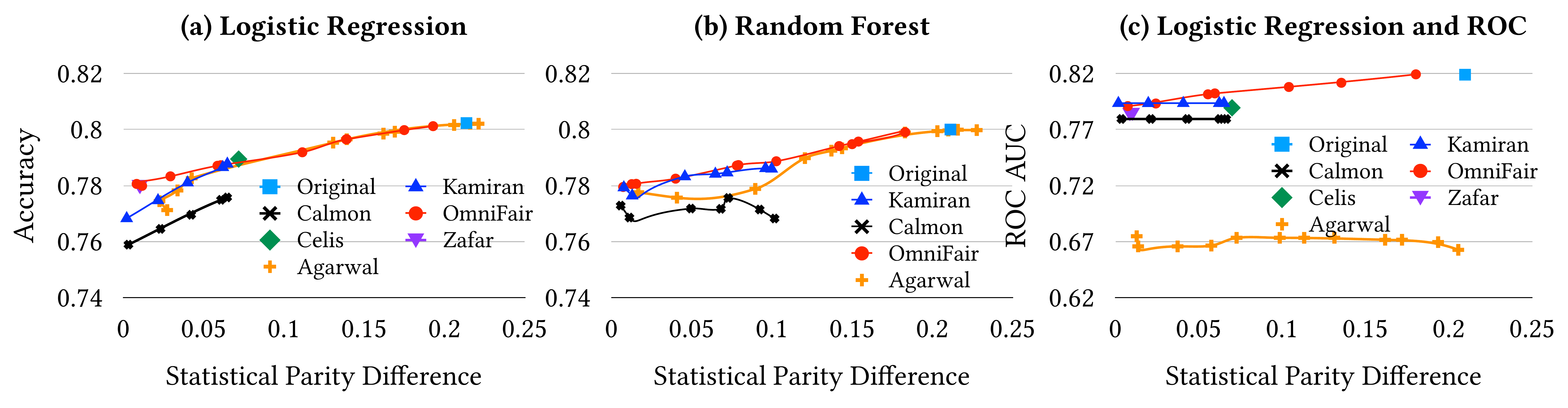}
\vspace{-5mm}
\end{minipage}
\hfill
\end{figure*}

\vspace{-1mm}
\subsection{Experiment Setup}
\label{sec:exp_setup}
\vspace{-1mm}
\stitle{Datasets.} We use four popular datasets in the algorithmic fairness literature that are known to have biases against minority groups. The detail of the datasets are shown in Table~\ref{tab:datasets}.

\stitle{ML Algorithms.} We include four ML algorithms, Logistic Regression (LR), Random Forest (RF), XGBoost (XGB), and \rev{Neural Networks (NN)}, to show that our system is model-agnostic. 
These are commonly used ML algorithms on structured data with completely different training processes --- 
LR and NN has an explicit loss function and is trained via gradient descent; RF and XGB~\cite{chen2016xgboost} has no explicit loss function and is trained by building decision trees.

\stitle{Baselines.} We compared with five baselines: two preprocessing approaches and three  in-processsing approaches, shown in Table~\ref{table:baselines}. \rev{We also compare with the fairness approach~\cite{thomas2019preventing} that relies on a specific new ML algorithm CMA-ES developed within.} 

\stitle{Hyperparameter Tuning.} ML algorithm performances are known to be influenced by hyperparameters (e.g., learning rate, tree-depth, etc). For algorithmic fairness work, there is usually an additional hyperparameter that controls the accuracy-fairness trade-off (e.g., $\lambda$ in \systemx). A prior work FairPrep~\cite{schelter2019fairprep} showed that many existing algorithmic fairness work only focused on tuning the hyperparameter that controls the accuracy-fairness trade-off, and neglected to tune the traditional ML algorithm hyperparameters, which can strongly impact the reported results. \rev{We randomly split each dataset to 60\% training, 20\% validation, and 20\% test. All hyperparameters (including both traditional ML hyperparameters that are specific to an ML algorithm and hyperparameters that are specific to fairness enforcement algorithms) are tuned using the validation set, and all reported results are on the unseen test set. All results reported are the average performance of 10 different random splits.}

\vspace{-3mm}
\subsection{Single Fairness Constraint}
In this section, we validate that our algorithm produces high quality results compared to other baselines using one fairness constraint given by a fairness specification $(g, f, \varepsilon)$. $g$ returns the groups that are defined on the sensitive attribute of each dataset.
For the fairness metric $f$, we experiment with statistical parity (SP) and false discovery rate (FDR). We choose these two metrics because SP is representative for the fairness metrics whose induced example weights are NOT parameterized by the model $\theta$, and FDR is representative for the fairness metrics whose induced example weights are parameterized by the model $\theta$ (c.f. Table~\ref{table:weights}). We will also include a customized fairness metric, as described in Example~\ref{ex:cutomized_metric}. 

\vspace{-4mm}
\subsubsection{Statistical parity as the fairness metric. \hspace{25mm}}

\stitle{Test accuracy comparison when $\varepsilon = 0.03$.} Instead of reporting absolute accuracy numbers, we report the accuracy drop of different methods compared with no fairness constraints in Table~\ref{table:accuracy}. For example, the first number $-1.2\%$ in Table~\ref{table:accuracy} means that, on the COMPAS dataset, the accuracy of the LR classifier obtained by \systemx when enforcing the SP constraint with $\varepsilon = 0.03$ is $-1.2\%$ less than the accuracy of the LR classifier obtained without any fairness constraint. We highlight several observations: \rev{(1) \systemx achieves \revnew{the} smallest accuracy drop in 8 out of 20 cases (4 datasets $\times$ 5 ML algorithms), and reduces the accuracy drop by up to $1 - 0.3/5.8 = 94.8\%$ compared with the second best method using the RF classifier on the LSAC dataset; (2) in the remaining 12/20 cases, \systemx is always a close second best method apart from CMA-ES; (3) CMA-ES is an evolution strategy for optimization used by \revnew{Thomas et al.}~\cite{thomas2019preventing} and it is not supported by all algorithms but \revnew{Thomas et al.}~\cite{thomas2019preventing}; }and (4) many existing methods either do not support some ML algorithms such as RF and XGBoost (i.e., NA(2)) or fail to satisfy the fairness constraint (i.e., NA(1)). In particular, Celis et al.~\cite{celis2019classification} is not able to generate a valid result at $\varepsilon=0.03$.

\rev{\stitle{Varying validation set size.} While our default validation set size is 20\% of the whole dataset, we conduct an ablation experiment to empirically observe the impact of validation set size on test accuracy and test bias. We vary the validation set size by using a subset of 20\% validation split, and use the same training and test set. 
As shown in Figure~\ref{fig:validation},  where the input fairness constraint is to enforce the SP difference between two groups to be within $\varepsilon = 0.03$, we can see that too small a validation set fails to generalize well (the test bias is bigger than 0.03); as we increase the validation set, the test set behavior stabilizes and its bias is close to 0.03. }




\stitle{Accuracy-Fairness Trade-off Comparison by Varying $\varepsilon$.} To further understand the accuracy-fairness trade-off, we compare all methods by varying $\varepsilon$. We only show the results for Adult dataset for space reasons, and refer readers to the full report~\cite{Full_Report} for other datasets, which reveal similar findings. Figure~\ref{fig:single_fairness_adult} shows that \systemx offers a much more flexible trade-off than existing methods in that it covers the entire span of different bias levels (i.e., the x-axis), while almost all baselines, except Agarwal et al~\cite{agarwal2018reductions}, do not cover the full x-axis. This is because all baselines, is spite of providing various knobs that can influence the final model, do not offer guarantees for how those knobs affect the trade-off. For example, for Zafar et al, while we tuned the provided hyperparameter extensively, we discovered that the best model is always the same for different $\varepsilon$ (hence, only one point in Figure~\ref{fig:single_fairness_adult}(a)). \systemx, on the other hand, has a hyperparameter $\lambda$ that monotonically controls the accuracy-fairness trade-off. Compared with  Agarwal et al~\cite{agarwal2018reductions}, which  offers a comparably flexible trade-off, \systemx achieves better accuracy, especially when $\varepsilon$ is small, as shown in Figure~\ref{fig:single_fairness_adult}. \rev{Given that the adult dataset has imbalanced labels (76\% negative), we also report ROC AUC in Figure~\ref{fig:single_fairness_adult}(c): while Agarwal et al has a low ROC AUC on adult dataset, \systemx is able to achieve both high accuracy and high ROC AUC while satisfying fairness constraints.}

\vspace{-2mm}
\begin{figure*}[ht]

\noindent\begin{minipage}[t]{0.24\textwidth}

\centering
\includegraphics[width=\linewidth]{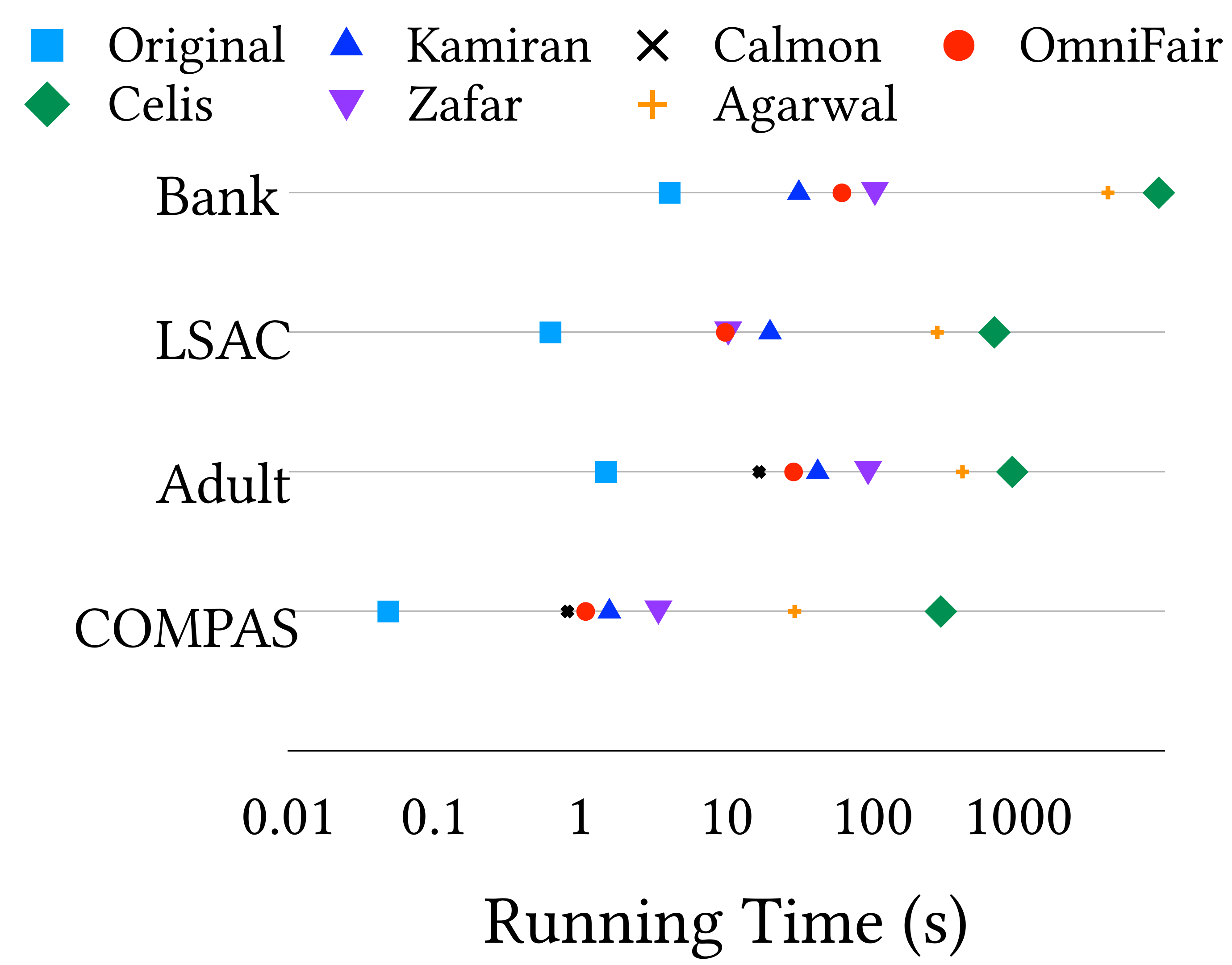}
\vspace{-7mm}\caption{\small Running time comparison under SP constraint and LR classifier.}
\label{fig:exp_efficiency_sp}
\vspace{-5mm}
\end{minipage}
\hfill
\begin{minipage}[t]{0.24\textwidth}
\noindent
\centering
\includegraphics[width=\linewidth]{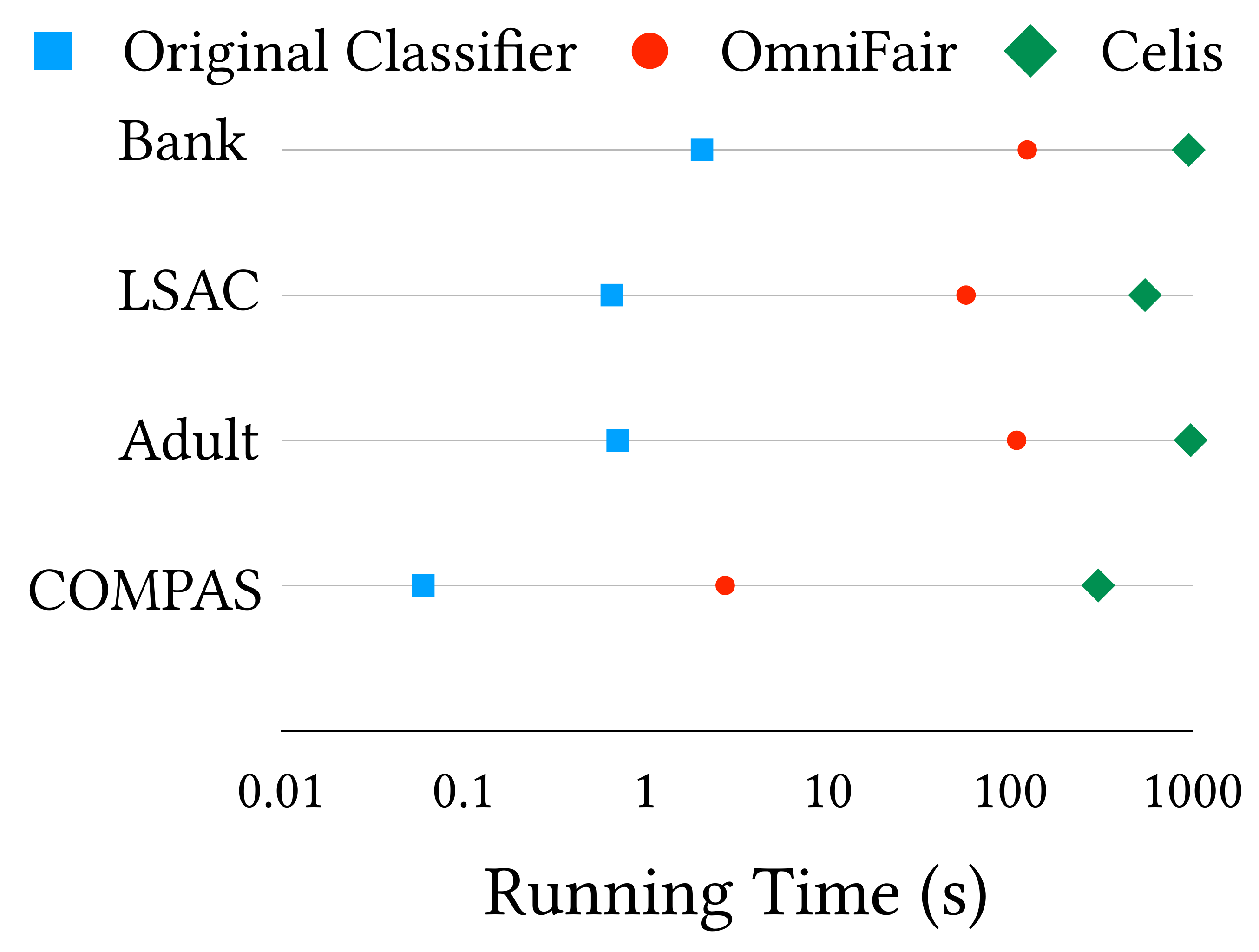}
\vspace{-7mm}\caption{\small Running time comparison under FDR constraint and LR classifier.}
\label{fig:exp_efficiency_fdr}
\vspace{-5mm}
\end{minipage}
\hfill
\begin{minipage}[t]{0.25\textwidth}
\centering
\includegraphics[width=\linewidth]{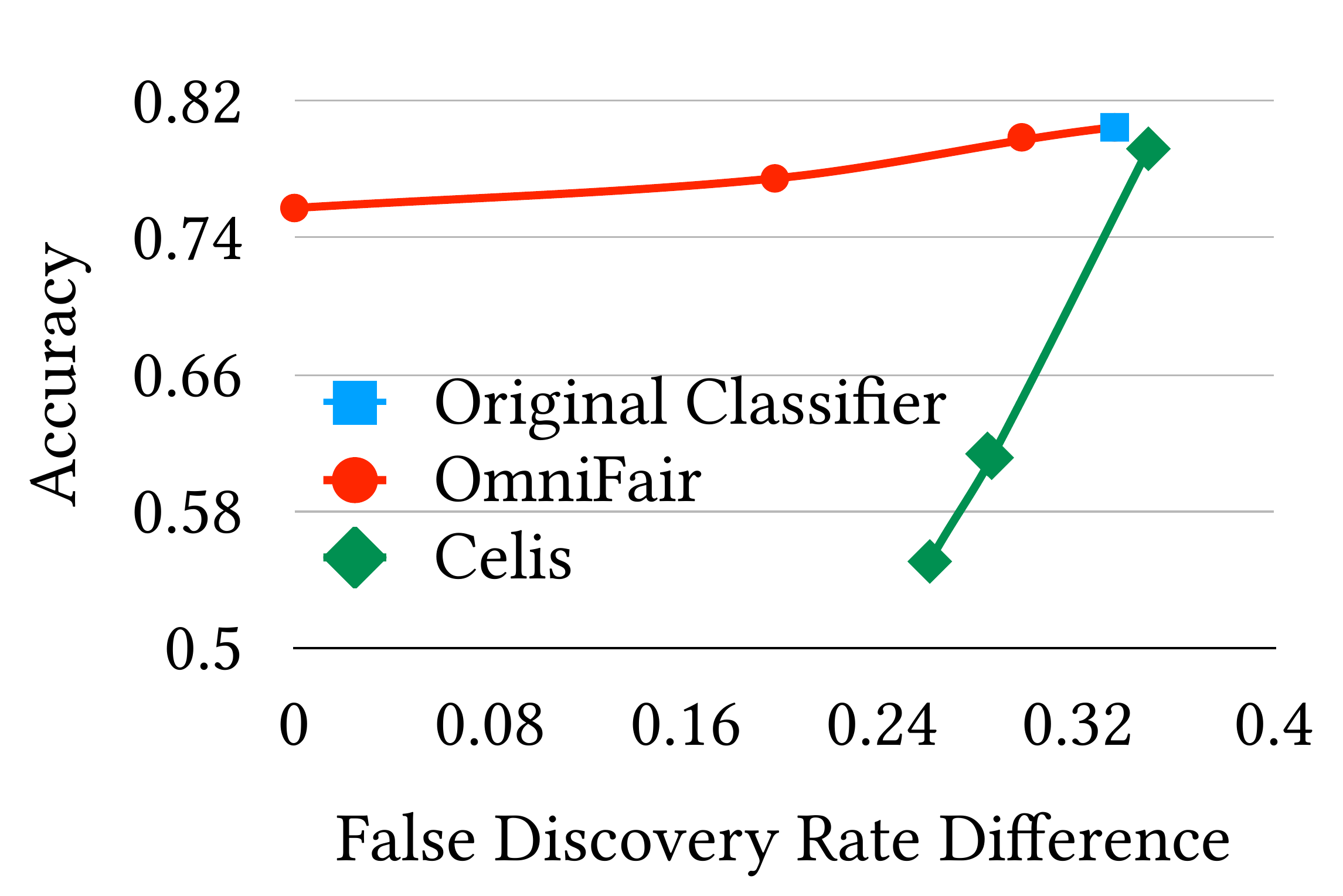}
\vspace{-7mm}\caption{\small The accuracy-fairness trade-off under FDR constraint and LR classifier on Adult dataset.} 
\label{fig:single_fairness_fdr_adult}
\vspace{-5mm}
\end{minipage}
\hfill
\begin{minipage}[t]{0.24\textwidth}
\centering
\includegraphics[width=\linewidth]{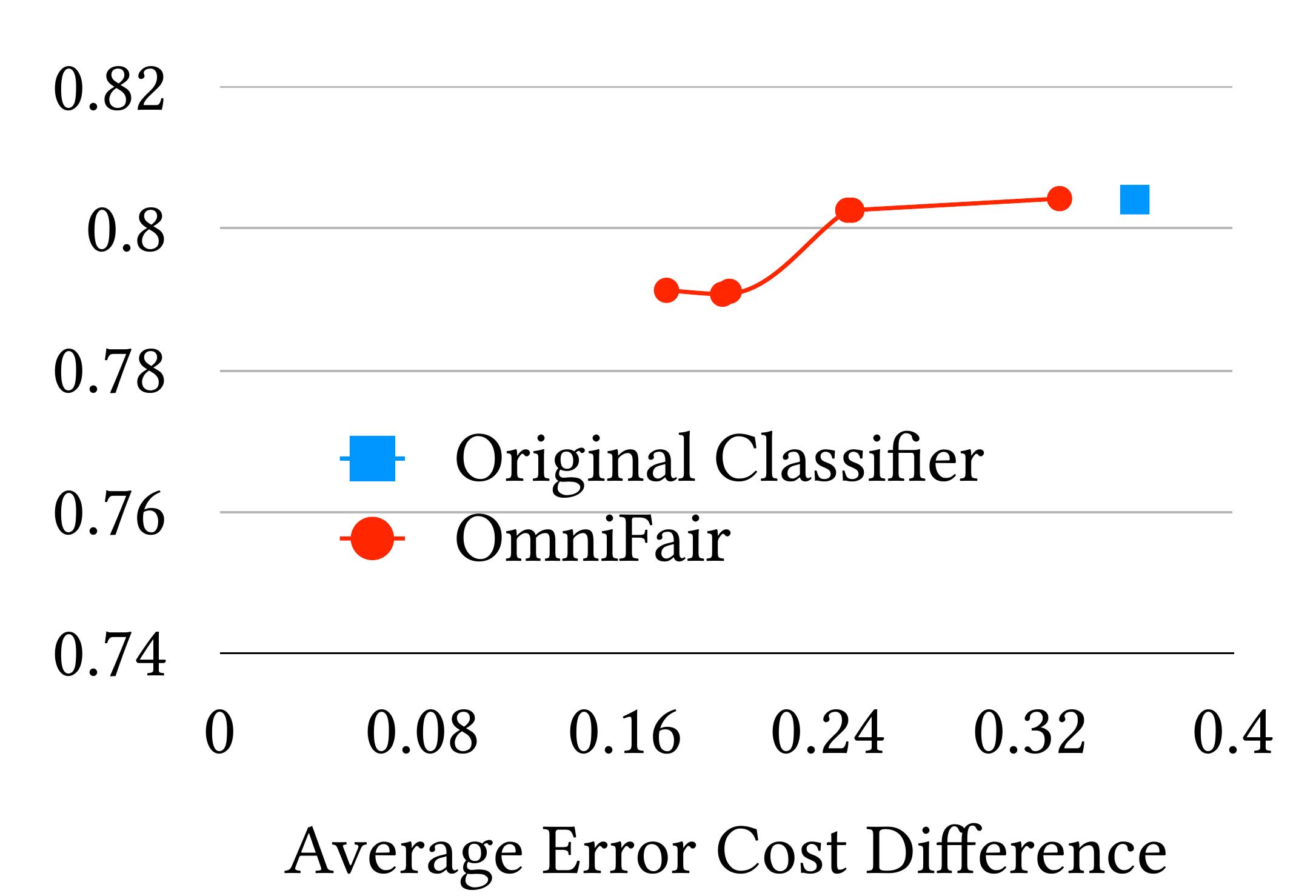}
\vspace{-7mm}\caption{\small The accuracy-fairness trade-off under AEC constraint and LR classifier on Adult dataset.} 
\label{fig:single_fairness_aec_adult}
\end{minipage}
\end{figure*}

\setlength{\tabcolsep}{1pt}

\begin{figure*}[ht]
\noindent\begin{minipage}[t]{0.25\textwidth}
\centering
\captionsetup{type=table} 
\centering
\small
\caption{\small The speed up for Warm Start under LR. Time is shown in seconds}
\vspace{-3mm}
\label{table:warmstart}
\scalebox{0.8}{
\begin{tabular}{|l|r|r|r|}
\hline
Dataset & No Warm Start (s) & Warm Start (s) & SpeedUp \\ \hline
Compas & 1.1 & 0.8 & 1.4$\times$ \\ \hline
Adult & 28.7 & 15.9 & 1.8$\times$ \\ \hline
LSAC & 9.8 & 8.2 & 1.2$\times$ \\ \hline
Bank & 25.6 & 7.5 & 3.4$\times$ \\ \hline   
\end{tabular}}
\vspace{-5mm}
\end{minipage}
\hfill
\begin{minipage}[t]{0.33\textwidth}
\centering
\vspace{-4mm}\caption{\small Enforcing SP for all three groups}
\includegraphics[width=0.8\linewidth]{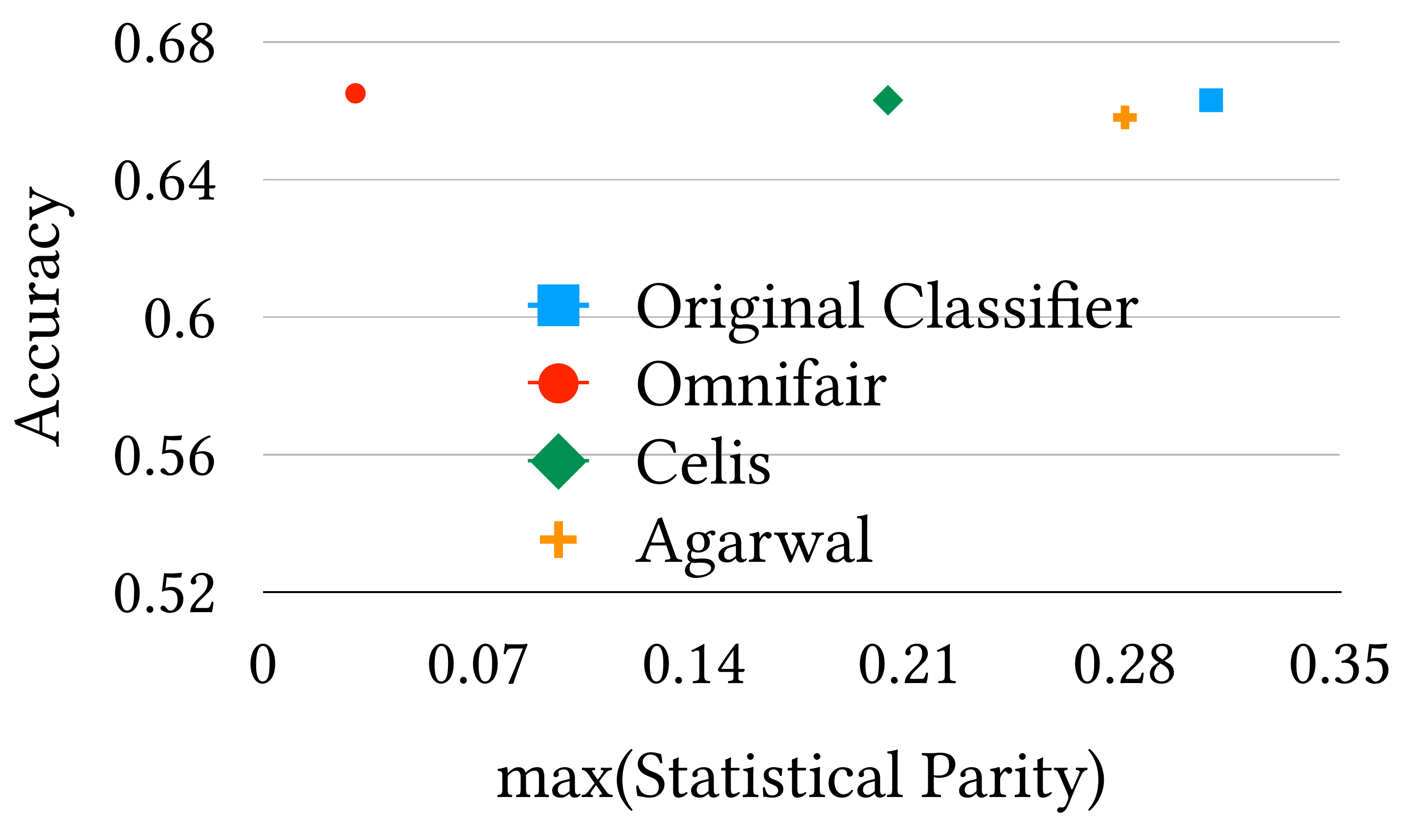}
\vspace{-5mm}
\label{fig:multi_group_cmp}
\vspace{-5mm}
\end{minipage}
\hfill
\setlength{\tabcolsep}{2pt}
\begin{minipage}[t]{0.20\textwidth}
\captionsetup{type=table} 
\centering
\small
\caption{\small Enforcing SP and FNR}
\vspace{-3mm}
\label{table:multiple_metric}
\scalebox{0.8}{
\begin{tabular}{|r|r|r|r|}
\hline
$\varepsilon$ & Accuracy & SP & FNR \\ \hline
Baseline & 63.9\% & 0.233 & 0.180 \\ \hline
$0.01$ & N/A & N/A & N/A \\ \hline
$0.02$ & N/A & N/A & N/A \\ \hline
$0.03$ & 63.6\% & 0.03 & 0.044 \\ \hline
$0.04$ & 63.4\% & 0.016 & 0.035 \\ \hline
$0.05$ & 63.3\% & 0.028 & 0.007 \\ \hline
$0.06$ & 62.7\% & 0.057 & 0.032 \\ \hline
\end{tabular}}
\vspace{-5mm}
\end{minipage}
\hfill
\begin{minipage}[t]{0.20\textwidth}
\captionsetup{type=table} 
\centering
\small
\caption{\small Grid vs hill-climbing (HC).}
\vspace{-3mm}
\label{table:efficiency}
\centering
\scalebox{0.8}{
\begin{tabular}{|l|l|l|r|r|}
\hline
$\varepsilon$ & Grid& HC & Grid Time & HC Time \\ \hline
$0.01$ & No & No & 96s & 9s\\ \hline
$0.02$ & No & No & 96s  & 8s\\ \hline
$0.03$ & No & Yes & 96s  & 7s\\ \hline
$0.04$ & Yes & Yes & 96s  & 7s\\ \hline
$0.05$ & Yes & Yes & 96s  & 7s\\ \hline
$0.06$ & Yes & Yes & 96s  & 7s\\ \hline

\end{tabular}
}
\end{minipage}
\end{figure*}
\stitle{Efficiency Comparison.} Figure~\ref{fig:exp_efficiency_sp} shows the comparison between the running time of different algorithms on Adult, COMPAS and LSAC datasets for statistical parity with LR. We only show the result for logistic regression here because it is supported by all baselines. The results for RF and XGB are similar, and could be found in our full report~\cite{Full_Report}.
\rev{Compared with existing pre-processing methods, OmniFair achieves comparable running time. However, OmniFair supports a much wider set of constraints, while pre-processing methods only support SP. Compared with existing in-processing methods, OmniFair is not only faster (from $1.05\times$ to $270\times$), but also achieves better accuracy-fairness trade-off (c.f. Figure~\ref{fig:single_fairness_adult}).}
         This is because in-processing methods such as Zafar et al. and Celis et al. require nontrivial modifications to ML training procedures. \rev{Compared with Agarwal et al., which also assumes black-box ML algorithms, \systemx is about 10X faster due to our efficient hyperparameter tuning using the monotonocity property while Agarwal et al. needs to solve a difficult saddle point finding problem. }

         \rev{We also implement a warm-start optimization for the LR algorithm (which is also applicable to NN) by using parameter values learned in a previous invocation of $\mathcal{A}$ as the initialization of parameter values in the next invocation of $\mathcal{A}$.  As shown in Table~\ref{table:warmstart}, warm-start leads to an 1.2$\times$ to 3.4$\times$ speed up on different datasets.}
 
 
\vspace{-4mm}
\subsubsection{False Discovery Rate and Customized Metric \hspace{16mm}}
\stitle{Accuracy-Fairness Trade-off Comparison by Varying $\varepsilon$.} Similar to SP, we do the analysis of the trade-off between accuracy and fairness by varying $\varepsilon$. We show that even for metrics like false discovery rate where the weight is a function of $\theta$, we are able to use our algorithm to achieve fairness. Among the five baselines, only Celis is able to support this. As shown in Figure~\ref{fig:single_fairness_fdr_adult}, our algorithm is able to reduce the false discovery rate difference while having little accuracy drop, and significantly outperform Celis. We show that for a customized metric error cost (AEC) as described in \S~\ref{sec:3:custom}, we are able to use our algorithm to reduce the bias while keeping similar accuracy to the baseline, as shown in Figure~\ref{fig:single_fairness_aec_adult}.

\stitle{Efficiency Comparison.} Figure~\ref{fig:exp_efficiency_fdr} shows the comparison between running time of different algorithms on Adult, COMPAS and LSAC datasets for false discovery rate with LR. We can see that for false discovery rate, we are $9\times$ to $150\times$ faster than Celis et. al., the only algorithm we found that could support false discovery rate.

\vspace{-3mm}
\subsection{Multiple Fairness Constraints}\label{sec:exp_multi}
We validate our algorithm when there are multiple fairness constraints. As discussed before, one fairness specification $(g, f, \varepsilon)$ can already generate multiple constraints if there are more than two groups in $g(D)$. Users could also use multiple fairness specifications to specify multiple constraints under different fairness metrics. We evaluate both cases. We also compare our algorithm with the naive grid search for hyperparameter tuning. 

\vspace{-1.5mm}
\stitle{Multiple groups under one fairness metric.} 
We train LR classifier on the COMPAS dataset. However, instead of considering only two groups: Black and White, we consider three groups: Black (B), White (W) and Hispanic (H). We want to minimize the statistical parity difference between any two groups. 

All baseline algorithms do not support this by default. We went through the code for all algorithms, and managed to adapt Celis et al. and Agarwal et al.'s code to support this.

The comparison is shown in Figure~\ref{fig:multi_group_cmp}. The x-axis is $SP_{max}$, the maximum value of SP difference among three groups, i.e., $SP_{max} = max(SP_{BW}, SP_{BH}, SP_{HW})$ and y-axis is accuracy. 
We can see that, although theoretically Celis and Agarwal can support multiple groups, they fail to reduce the SP difference among all three groups (the $SP_{max}$ is still greater than 0.20). On the other hand, \systemx can reduce $SP_{max}$ to 0.03, while keeping high accuracy.





\vspace{-1.5mm}
\stitle{Multiple fairness metrics.} 
We run the experiments on COMPAS dataset considering two fairness metrics: SP and FNR. 

Theoretically, this could be supported by Celis and Agarwal;  however, we failed to modify their code to support multiple metrics after spending two days on it as constraints are hard-coded into the optimization procedure.
In contrast, \systemx supports multiple metrics without any additional coding efforts.

As we can see in Table~\ref{table:multiple_metric}, for $\varepsilon=0.01$ and $\varepsilon=0.02$, our algorithm is not able to find a result. This is because we cannot find a set of weights such that $\varepsilon$ is satisfied for both constraints in the validation set. In these cases, neither could grid search return a feasible answer. However, for $\varepsilon >= 0.03$, we can see that the fairness difference for both fairness constraints drop by one order of magnitude, and the accuracy loss is minimum ($< 1\%$). 

\vspace{-1.5mm}
\stitle{Efficiency Optimization.} We compare the hill-climbing algorithm with grid search in terms of efficiency and the ability to discover feasible solutions. 
We run the experiments on the COMPAS dataset and consider two fairness constraints: statistical parity and false negative rate with different degrees of disparity allowance $\varepsilon$. As we can see in Table~\ref{table:efficiency}, when Grid Search finds a feasible solution, the hill-climbing algorithm can always find one as well. Also, the hill-climbing algorithm runs around $10\times$ than grid search.

\vspace{-3mm}
\section{Conclusion and Future Work} \label{sec:conclusion}
In this paper, we propose \systemx that allows users to declaratively specify group fairness constraints in ML. \systemx is versatile in that it supports all major group fairness constraints, including customized ones and multiple constraints simultaneously, and it is model-agnostic. The algorithms in \systemx maximize for accuracy while satisfying given fairness constraints. The algorithm designs are optimized based on monotonocity properties for  accuracy and fairness unique in our system. Experimental results demonstrate that \systemx supports more fairness metrics than existing methods, and can achieve better accuracy-fairness trade-off without sacrificing efficiency. \rev{Future work includes (1) improving the scalability of OmniFair, which could potentially be addressed by parallel model training and  using a smaller sample training set to quickly prune certain $\lambda$ values.; and (2) providing theoretical guarantees regarding constraint satisfiability on unseen test sets.}

\balance

\break
\bibliographystyle{abbrv}
\bibliography{ref}
\newpage
\appendix

\section{Expressing Fairness Metrics}

\stitle{Expressing MR.} Consider MR as the fairness metric where the goal is to ensure demographic groups have the same misclassification rates. This fairness metric is equivalent to equalizing {\em accuracy} across different groups, i.e., 
\begin{equation}
 \at{f}(h,g) = \frac{1}{\vert g \vert} \sum_{i \in g} \mathbbm{1} (h(x_i)=y_i)   
\end{equation}
\stitle{Expressing FOR.}
As shown in ~\Cref{equation:pp_pos}, under FOR, for each group, the fairness metric is computed as follows:
\begin{equation}
\footnotesize
\begin{split}
    &f(h,g) = Pr(y=1 \vert  h(x)=0) = \frac{\vert \{i : i \in g, h(x_i) = 0, y_i = 1\}\vert}{\vert \{i : i \in g, h(x_i) = 0\}\vert} \\
    &= 1 - \frac{1}{{\vert \{i : i \in g, h(x_i) = 0\}\vert} } \sum_{ \{i : i \in g, h(x_i) = 0\}}  \mathbbm{1} (h(x_i)=y_i) \\
    &= 1 - \frac{1}{{\vert \{i : i \in g, h(x_i) = 0\}\vert} } \sum_{ \{i : i \in g, y_i = 0\}}  \mathbbm{1} (h(x_i)=y_i) 
\end{split}
\end{equation}
The above derivation shows that, for data points with $y_i=0$, the coefficients are $c_i = -\frac{1}{{\vert \{i : i \in g, h(x_i) = 0\}\vert} }$. For data points with $y_i = 1$, the coefficients are $c_i = 0$. Also, we have $c_0=1$.

\stitle{Expressing Average Cost of Error.} In  Example~\ref{runningEx2}, a false negative is to predict an individual who is likely to re-offend to be low risk, could damage the community if the criminal indeed re-offends in the future. On the other hand, a false positive is to predict an individual who is unlikely to re-offend to be high risk, could keep the criminal in the jail for an unnecessarily long time. The cost for making these two types of errors are different. 

Suppose a user is interested in ensuring that the average costs of error between the \attrib{African American} group and the \attrib{Caucasian} group are similar, under some cost $C_{fp}$ of making a false positive error and some cost $C_{fn}$ of making a false negative error. To this end, the user defines the following customized fairness metric function: 

\begin{equation}
\footnotesize
\begin{split}
 & f(h,g)  = \frac{C_{fp} * \vert h(x)=1, y=0 \vert + C_{fn} * \vert h(x)=0, y=1\vert}{\vert g \vert } \\
 & =  \frac{  C_{fp}* \sum\limits_{\{i : i \in g, y = 0\}} (1-\mathbbm{1} (h(x_i)=y_i))}{\vert g \vert } + \frac{ C_{fn}*  \sum\limits_{ \{i : i \in g, y = 1\}} (1-\mathbbm{1} (h(x_i)=y_i)) }{\vert g \vert} \nonumber
\end{split}
\end{equation}
where $C_{fp}$ is the cost to make a false positive error, $C_{fn}$ is the cost to make a false negative error and $ \vert g \vert$ is the total number of samples in group $g$. $C_{fp}$ and $C_{fn}$ can be customized by the user. We want this metric to have same or similar value among different groups.

Therefore, for data points with label $y_i = 0$ (resp. $y_i=1$), the coefficient is $c_i = -C_{fp}/\vert g \vert$ (resp. $c_i = -C_{fn}/\vert g \vert$). Also, we have $c_0 = \big (C_{fp}*\vert \{i : i \in g, y_i = 0\} \vert +C_{fn}*\vert \{i : i \in g, y_i = 1\} \vert \big) / \vert g \vert$.

\revnew{
\section{Proof for Lemma 1}

\medskip
\begin{proof}

We organize our proof in the following two steps, which corresponds the two parts in Lemma 1. The first step is directly following the Lagrangian dual function and its results, as discussed in Chapter 5 of Boyd et al.~\cite{boyd2004convex}.

\medskip
\stitle{Step 1: We rewrite~\Cref{equ:original}, show its Lagrangian dual gives an upper bound to~\Cref{equ:original}, and leverages the strong duality.}

We rewrite~\Cref{equ:original} into an equivalent~\Cref{equ:both_direc} by re-writing its constraint with the absolute term into two constraints: 
\begin{equation}~\label{equ:both_direc}
\begin{split}
       \max_{\theta} \quad & AP(\theta)  \\
        \text{s.t.} \quad  &\varepsilon - FP(\theta) \geq 0  \\
        \quad &\varepsilon + FP(\theta) \geq 0 \\
    \end{split}
\end{equation}

Following the standard Lagrangian duality formulation, function L is defined as  
\begin{equation}
\begin{split}
L(\theta,\lambda_1,\lambda_2) =&  AP(\theta) + \lambda_1 (\varepsilon - FP(\theta)) + \lambda_2 (\varepsilon + FP(\theta) ) \\
=& AP(\theta) + (\lambda_2 - \lambda_1) FP(\theta) + (\lambda_1 + \lambda_2) \varepsilon
\end{split}
\end{equation}

and the Lagrangian dual function $h(\lambda_1,\lambda_2)$ is defined as: 
\begin{equation*}\label{equ:h_L}
\begin{split}
h(\lambda_1,\lambda_2) = &\max_{\theta} L(\theta,\lambda_1,\lambda_2)
\end{split}
\end{equation*}
In other words, the Lagrangian dual function $h(\lambda_1,\lambda_2)$ finds the optimal value to $L(\theta,\lambda_1,\lambda_2)$ for a given $\lambda_1$ and $\lambda_2$. We next show that if $\theta^*$ is an optimal solution to~\Cref{equ:original}, for all $\lambda_1, \lambda_2 > 0:$ $h(\lambda_1,\lambda_2) \geq AP(\theta^*)$.

Given that $\theta^*$ is also a feasible solution to~\Cref{equ:original}, we have $\varepsilon -FP(\theta^*) \geq 0 $ and $\varepsilon  + FP(\theta^*) \geq 0$, so $\lambda_1 (\varepsilon - FP(\theta^*)) + \lambda_2 (\varepsilon + FP(\theta^*)) \geq 0$, for $\lambda_1 > 0$ and $\lambda_2 > 0$. We thus have $L(\theta^*,\lambda_1,\lambda_2) \geq AP(\theta^*)$. We also have  $h(\lambda_1,\lambda_2) = \max_{\theta} L(\theta,\lambda_1,\lambda_2) \geq L(\theta^*,\lambda_1,\lambda_2)$ by definition.

Hence, we have for $\lambda_1 > 0$ and $\lambda_2 > 0$, $h(\lambda_1,\lambda_2) \geq AP(\theta^*)$, in other worlds, for $\lambda_1 > 0$ and $\lambda_2 > 0$, the Lagrangian dual function $h(\lambda_1,\lambda_2)$ provides an upper bound to the original problem.  

Among all the upper bounds, we want to find the tightest (smallest) upper bound, so the Lagrangian dual optimization problem is defined as:
\begin{equation}\label{equ:dual}
    \begin{split}
       \min_{\lambda_1, \lambda_2} \quad &h(\lambda_1,\lambda_2)  \\
        \text{s.t.} \quad &\lambda_1,\lambda_2 > 0\\
    \end{split}
\end{equation}

In other words, the Lagrangian dual optimization problem ~\Cref{equ:dual} finds the minimal value of $h(\lambda_1,\lambda_2)$ among all possible positive  $\lambda_1$ and  $\lambda_2$.

We have shown that the optimal value to the dual problem ~\Cref{equ:dual} is always greater than or equal to the the optimal value to the primal problem ~\Cref{equ:both_direc}, namely, $\min_{\lambda_1 > 0,\lambda_2>0} h(\lambda_1,\lambda_2) >= AP(\theta^*)$
The difference between the optimal values to these two problems are referred to as the duality gap, and the fact that the optimal value to the dual problem ~\Cref{equ:dual} always provides an upper bound to the optimal value to the primal problem ~\Cref{equ:both_direc} (in the case of maximization) is a concept referred to as the \textit{weak duality}~\cite{boyd2004convex}.





\textit{Strong duality assumption}~\cite{boyd2004convex} states that optimizing the dual problem~\Cref{equ:dual} gives the exact same optimal solution to optimizing the primal problem~\Cref{equ:both_direc}.
In other words, we can find a specific $\lambda_1^* > 0$ and $\lambda_2^* > 0$, such that the duality gap is zero, i.e.,  $h(\lambda_1^*,\lambda_2^*) = AP(\theta^*)$, which means that $\theta^*$ also optimizes $h(\lambda_1^*,\lambda_2^*)$.



\medskip
\underline{\textit{Some discussions on strong duality assumption}.} 
\begin{itemize}
    \item There are multiple different sufficient conditions for satisfying the strong duality assumption, and this has been an important topic in optimization theory. For example, if the primal problem is a linear optimization problem, then strong duality is satisfied automatically. Another commonly used condition for satisfying strong duality is if the primal problem is a convex optimization problem and the Slater's condition holds (c.f. Chapter 5.2.3 in~\cite{boyd2004convex}), which  states that the primal problem has a  solution  such that the inequalities in the constraints are satisfied strictly. 
    
    \item Even when strong duality assumption fails to hold, the dual problem is still commonly used to find  good feasible solutions to the primal problem in practice. This is exactly \systemx's design innovation: while we cannot solve the constrained optimization problem in ~\Cref{equ:both_direc} in a model-agnostic way, we are able to solve ~\Cref{equ:h} in a model-agnostic way. 
\end{itemize}


\medskip
\stitle{Step 2: Proving the existence of $\lambda$:}
We have shown that instead of solving the original problem~\Cref{equ:original}, we can instead solve the Lagrangian dual optimization problem~\Cref{equ:h}, which gives an upper bound to the original problem and can be a good approximation. The bound is even tight when the strong duality holds.

Next we show that we can further simplify the dual function~\Cref{equ:h} with~\Cref{equ:reg2}, i.e., there exists $\lambda$ such that for $\tilde{\theta}$ that optimizes~\Cref{equ:h}, it also optimizes~\Cref{equ:reg2}.



We let $\lambda = \lambda_2 - \lambda_1$, where $\lambda_1 > 0$ and $\lambda_2 > 0$ are the two values that make $\tilde{\theta}$ the optimal solution to $h(\lambda_1, \lambda_2)$. 

We prove this by contradiction. Assume $\tilde{\theta}$ is not the optimal solution to ~\Cref{equ:reg2}. In other words, there exists $\hat{\theta} \neq \tilde{\theta}, \text{s.t.}, AP(\hat{\theta})+\lambda FP(\hat{\theta}) > AP(\tilde{\theta})+\lambda FP(\tilde{\theta})$. This means that: 
\begin{equation}
\begin{split}
    AP(\hat{\theta}) + (\lambda_2 - \lambda_1) FP(\hat{\theta}) &> AP(\tilde{\theta}) + (\lambda_2 - \lambda_1) FP(\tilde{\theta}) \\
    \Longrightarrow AP(\hat{\theta}) + (\lambda_2 - \lambda_1) FP(\hat{\theta})& + (\lambda_1 + \lambda_2) \varepsilon  \\
    >  AP(\tilde{\theta}) &+ (\lambda_2 - \lambda_1) FP(\tilde{\theta}) + (\lambda_1 + \lambda_2) \varepsilon \\
\end{split}
\end{equation}

The above equation directly contradicts our already known fact that $\tilde{\theta}$ is the optimal solution to $h(\lambda_1, \lambda_2) = \max_{\theta} AP(\theta) + (\lambda_2 - \lambda_1) FP(\theta) + (\lambda_1 + \lambda_2) \varepsilon$.

Under strong duality assumption, in Step 1 we have shown that exists $\lambda_1^*, \lambda_2^*>0$, such that $h(\lambda_1^*,\lambda_2^*) = L(\theta^*,\lambda_1^*,\lambda_2^*)$, i.e., $\theta^*$ also optimizes~\Cref{equ:h} with $\lambda_1^*$ and $\lambda_2^*$. Therefore, there  exists $\lambda = \lambda_2^* - \lambda_1^*$, such that $\theta^*$ also optimizes~\Cref{equ:reg2}.

\end{proof}
}

\section{Proof for Lemma 2}
\begin{proof} 

Given that $\theta_1$ and $\theta_2$ are two optimal solutions to ~\Cref{equ:lambda_1} and~\Cref{equ:lambda_2}, respectively. \revnew{Here we assume that the optimal solutions exist}. We have 
\begin{equation}
\small
    \label{equ:lambda1}
    AP(\theta_1) + \lambda_1 FP(\theta_1) \geq  AP(\theta_2) + \lambda_1 FP(\theta_2)
\end{equation}
\begin{equation}
\small
    \label{equ:lambda2}
    AP(\theta_2) + \lambda_2 FP(\theta_2) \geq  AP(\theta_1) + \lambda_2 FP(\theta_1)
\end{equation}
Adding \Cref{equ:lambda1} and~\Cref{equ:lambda2}, we have
\begin{equation}
\small
\label{equ:lambda_1_lambda_sum}
    \lambda_1 FP(\theta_1) + \lambda_2 FP(\theta_2) \geq  \lambda_1 FP(\theta_2) + \lambda_2 FP(\theta_1)
\end{equation}
By rearranging terms in~\Cref{equ:lambda_1_lambda_sum}, we get
\begin{equation}
\small
    (\lambda_2 - \lambda_1)(FP(\theta_2) -FP(\theta_1) ) \geq 0
\end{equation}
Given  $\lambda_2 > \lambda_1$, we thus have $FP(\theta_1) \leq FP(\theta_2)$ in ~\Cref{equ:fp_monotonic}.

If $\lambda_2 > \lambda_1 \geq 0$, by shuffling the terms in ~\Cref{equ:lambda1}, we can have $AP(\theta_1)  \geq  AP(\theta_2) + \lambda_1 (FP(\theta_2)-FP(\theta_1)) \geq AP(\theta_2) $, which proves~\Cref{equ:ap_monotonic_pos}.

Similarly, If $0 \geq \lambda_2 > \lambda_1$, by shuffling the terms in ~\Cref{equ:lambda2}, we can have $AP(\theta_2)  \geq  AP(\theta_1) + \lambda_2 (FP(\theta_1)-FP(\theta_2)) \geq AP(\theta_1)$, which  proves~\Cref{equ:ap_monotonic_neg}.
\end{proof}

\revnew{
\section{Proof for Lemma 3}

\medskip
\begin{proof}

The proof of Lemma 3 is similar to the proof of Lemma 1, except that we now have multiple constraints and have a vector $\Lambda$ instead on single $\lambda$. For completeness, we detail the proof of Lemma 3 via similar two steps in the following.


\medskip
\stitle{Step 1: We rewrite~\Cref{equ:original_multi}, show its Lagrangian dual gives an upper bound to~\Cref{equ:original_multi}, and leverage the strong duality.}

We rewrite~\Cref{equ:original_multi} into an equivalent~\Cref{equ:both_direc_multi} by re-writing its constraint with the absolute term into two constraints: 
\begin{equation}~\label{equ:both_direc_multi}
\begin{split}
       \max_{\theta} \quad & AP(\theta)  \\
        \text{s.t.} \quad &\varepsilon - FP_i(\theta) \geq 0 \quad \forall i \in \{1..k\}\\
        \quad &\varepsilon + FP_i(\theta) \geq 0  \quad  \forall i \in \{1..k\}\\
    \end{split}
\end{equation}

Following the standard Lagrangian duality formulation, function L is defined as  
\begin{equation}
\begin{split}
L(\theta,\Lambda_1,\Lambda_2) =&  AP(\theta) + \sum_{i=1}^k\lambda_{1i} (\varepsilon - FP_i(\theta)) + \sum_{i=1}^k\lambda_{2i} (\varepsilon + FP_i(\theta)  ) \\
=& AP(\theta) + \sum_{i=1}^k(\lambda_{2i} - \lambda_{1i}) FP_i(\theta) + \sum_{i=1}^k(\lambda_{1i} + \lambda_{2i}) \varepsilon
\end{split}
\end{equation}

and the Lagrangian dual function $h(\Lambda_1,\Lambda_2)$ is:
\begin{equation}
\tag{\ref{equ:h_multi}}
\begin{split}
h(\Lambda_1,\Lambda_2) = &\max_{\theta} L(\theta,\Lambda_1,\Lambda_2)
\end{split}
\end{equation}
In other words, the Lagrangian dual function $h(\Lambda_1,\Lambda_2)$ finds the optimal value to $L(\theta,\Lambda_1,\Lambda_2)$ for a given $\Lambda_1$ and $\Lambda_2$. We next show that if $\theta^*$ is an optimal solution to~\Cref{equ:original_multi}, for all $\Lambda_1, \Lambda_2 > 0:$ $h(\Lambda_1,\Lambda_2) \geq AP(\theta^*)$.

Given that $\theta^*$ is also a feasible solution to~\Cref{equ:original_multi}, we have $\varepsilon - FP_i(\theta^*) \geq 0 $ and $\varepsilon + FP_i(\theta^*) \geq 0$, so $\sum_{i=1}^k \lambda_{1i} (\varepsilon - FP_i(\theta^*)) + \sum_{i=1}^k\lambda_{2i} (\varepsilon + FP_i(\theta^*)) \geq 0$, for $\Lambda_1 > 0$ and $\Lambda_2 > 0$. We thus have $L(\theta^*,\Lambda_1,\Lambda_2) \geq AP(\theta^*)$. We also have  $h(\Lambda_1,\Lambda_2) = \max_{\theta} L(\theta,\Lambda_1,\Lambda_2) \geq L(\theta^*,\Lambda_1,\Lambda_2)$ by definition.

Hence, we have for $\Lambda_1 > 0$ and $\Lambda_2 > 0$, $h(\Lambda_1,\Lambda_2) \geq AP(\theta^*)$. In other words, for $\Lambda_1 > 0$ and $\Lambda_2 > 0$, the Lagrangian dual function $h(\Lambda_1,\Lambda_2)$ provides an upper bound to the original problem.  

Among all the upper bounds, we want to find the tightest (smallest) upper bound, so the Lagrangian dual optimization problem is defined as:
\begin{equation}\label{equ:dual_multi}
    \begin{split}
       \min_{\Lambda_1, \Lambda_2} \quad &h(\Lambda_1,\Lambda_2)  \\
        \text{s.t.} \quad &\Lambda_1,\Lambda_2 > 0\\
    \end{split}
\end{equation}

In other words, the Lagrangian dual optimization problem ~\Cref{equ:dual_multi} finds the minimal value of $h(\Lambda_1, \Lambda_2)$ among all possible positive  $\Lambda_1$ and  $\Lambda_2$.

What \textit{strong duality assumption}~\cite{boyd2004convex} states is that optimizing the dual problem~\Cref{equ:dual_multi} gives the exact same optimal solution to optimizing the primal problem~\Cref{equ:both_direc_multi}. That is, we can find a specific $\Lambda_1^* > 0$ and $\Lambda_2^* > 0$, such that the duality gap is zero, i.e.,  $h(\Lambda_1^*,\Lambda_2^*) = AP(\theta^*)$, which means that $\theta^*$ also optimizes $h(\Lambda_1^*,\Lambda_2^*)$.


\medskip
\stitle{Step 2: Proving the existence of $\Lambda$:}
We have shown that instead of solving the original problem~\Cref{equ:original_multi}, we can instead solve the Lagrangian dual optimization problem~\Cref{equ:h_multi}, which gives an upper bound to the original problem and can be a good approximation. The bound is even tight when the strong duality holds.

Next we show that we can further simplify the dual function~\Cref{equ:h_multi} with~\Cref{equation:fairness_opt_multiple}, i.e., there exists $\Lambda$ such that for $\tilde{\theta}$ that optimizes~\Cref{equ:h_multi}, it also optimizes~\Cref{equation:fairness_opt_multiple}.



We let $\Lambda = \Lambda_2 - \Lambda_1$, where $\Lambda_1 > 0$ and $\Lambda_2 > 0$ are the two values that make $\tilde{\theta}$ the optimal solution to $h(\Lambda_1, \Lambda_2)$. 

We prove this by contradiction. Assume $\tilde{\theta}$ is not the optimal solution to ~\Cref{equation:fairness_opt_multiple}. In other words, there exists $\hat{\theta} \neq \tilde{\theta}, \text{s.t.}, AP(\hat{\theta})+\sum_{i=1}^k \lambda_i FP_i(\hat{\theta}) > AP(\tilde{\theta})+\sum_{i=1}^k\lambda_i FP_i(\tilde{\theta})$. This means that: 
\begin{equation}
\begin{split}
    AP(\hat{\theta}) + \sum_{i=1}^k (\lambda_{2i} - \lambda_{1i}) FP_i(\hat{\theta}) &> AP(\tilde{\theta}) + \sum_{i=1}^k (\lambda_{2i} - \lambda_{1i}) FP_i(\tilde{\theta}) \\
    \Longrightarrow AP(\hat{\theta}) + \sum_{i=1}^k(\lambda_{2i} - \lambda_{1i}) FP_i(\hat{\theta})& +  \sum_{i=1}^k(\lambda_{1i} + \lambda_{2i}) \varepsilon  \\
    >  AP(\tilde{\theta}) + \sum_{i=1}^k(\lambda_{2i} &- \lambda_{1i}) FP_i(\tilde{\theta}) + \sum_{i=1}^k(\lambda_{1i} + \lambda_{2i}) \varepsilon \\
\end{split}
\end{equation}

The above equation directly contradicts our already known fact that $\tilde{\theta}$ is the optimal solution to $h(\Lambda_1, \Lambda_2) = \max_{\theta} AP(\theta) + \sum_{i=1}^k(\lambda_{2i} - \lambda_{1i})FP_i(\theta) + (\Lambda_1 + \Lambda_2) \varepsilon$.

Under strong duality assumption, in Step 1 we have shown that exists $\Lambda_1^*, \Lambda_2^*>0$, such that $h(\Lambda_1^*,\Lambda_2^*) = L(\theta^*,\Lambda_1^*,\Lambda_2^*)$, i.e., $\theta^*$ also optimizes~\Cref{equ:h_multi} with $\Lambda_1^*$ and $\Lambda_2^*$. Therefore, there exists $\Lambda = \Lambda_2^* - \Lambda_1^*$ such that $\theta^*$ also optimizes~\Cref{equation:fairness_opt_multiple}.

\end{proof}

}

\section{Additional experiment results}
Due to space limit, we are not able to show all results in the main paper, so we show the extra experimental results here in this section.

\subsection{Statistical parity as the fairness metric.}

\begin{figure}[h]
        \centering
\includegraphics[width=\linewidth]{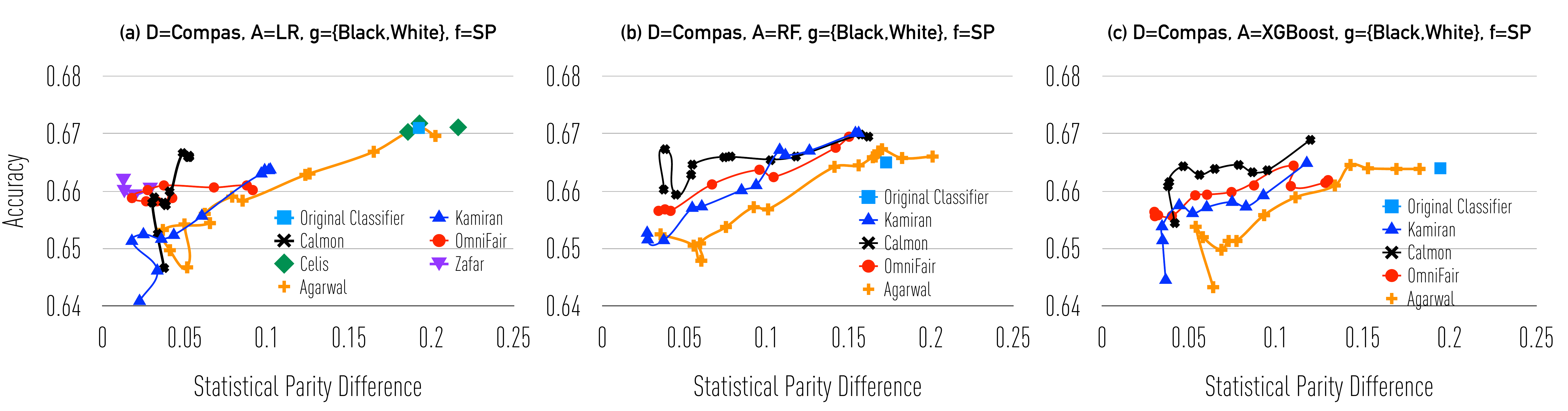}
\caption{\small The trade-off between fairness metric and accuracy with LR, RF and XGBoost on Compas} 
\label{fig:single_fairness_compas}
\end{figure}

\begin{figure}[h]
        \centering
\includegraphics[width=\linewidth]{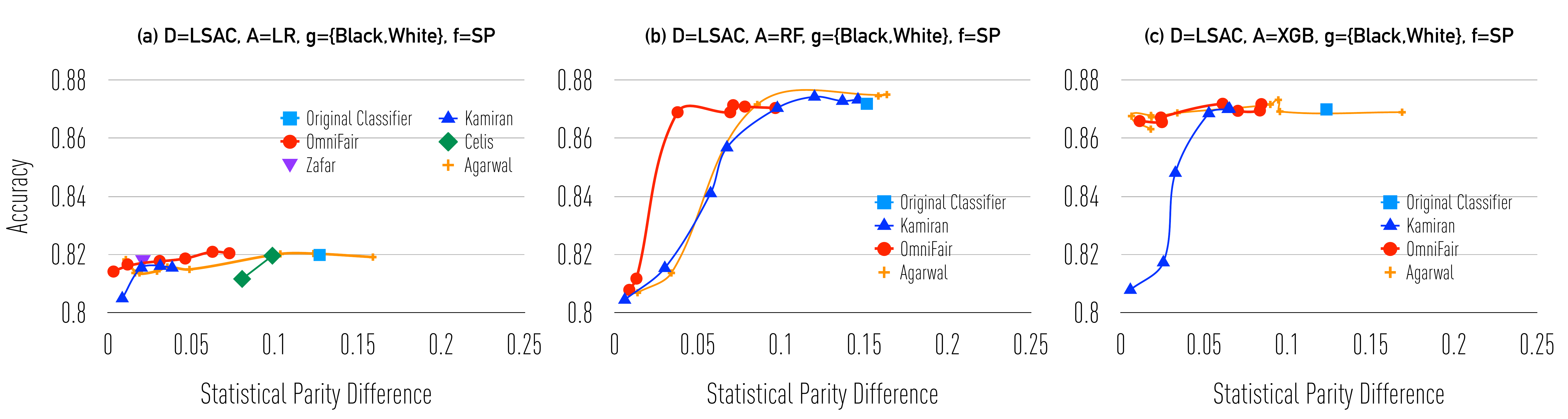}
\caption{\small The trade-off between fairness metric and accuracy with LR, RF and XGBoost on LSAC} 
\label{fig:single_fairness_lsac}
\end{figure}

\stitle{Accuracy-Fairness Trade-off Comparison by Varying $\varepsilon$.} To further understand the accuracy-fairness trade-off, we compare all methods by varying $\varepsilon$.  Figure~\ref{fig:single_fairness_compas} and~\ref{fig:single_fairness_lsac} shows the result for COMPAS dataset and LSAC dataset, respectively. The result is similar to the result in Section~\ref{sec:exp}. On both datasets, \systemx is able to cover the full x-axis, meaning that it can cover the entire span of the different bias levels. On COMPAS dataset, \systemx is the second best performing model, next to Calmon. On LSAC dataset, \systemx is the best performing model, with the highest accuracy, showing that \systemx is able to achieve good accuracy when getting certain fairness level. Notice that The Calmon model doesn't work for LSAC dataset, as it needs to specify a dataset specific parameter, and the authors only provided the parameter for Adult and COMPAS dataset.

\subsection{False Discovery Rate and Customized Metric }
\begin{figure}[h]
\centering
\includegraphics[width=0.6\linewidth]{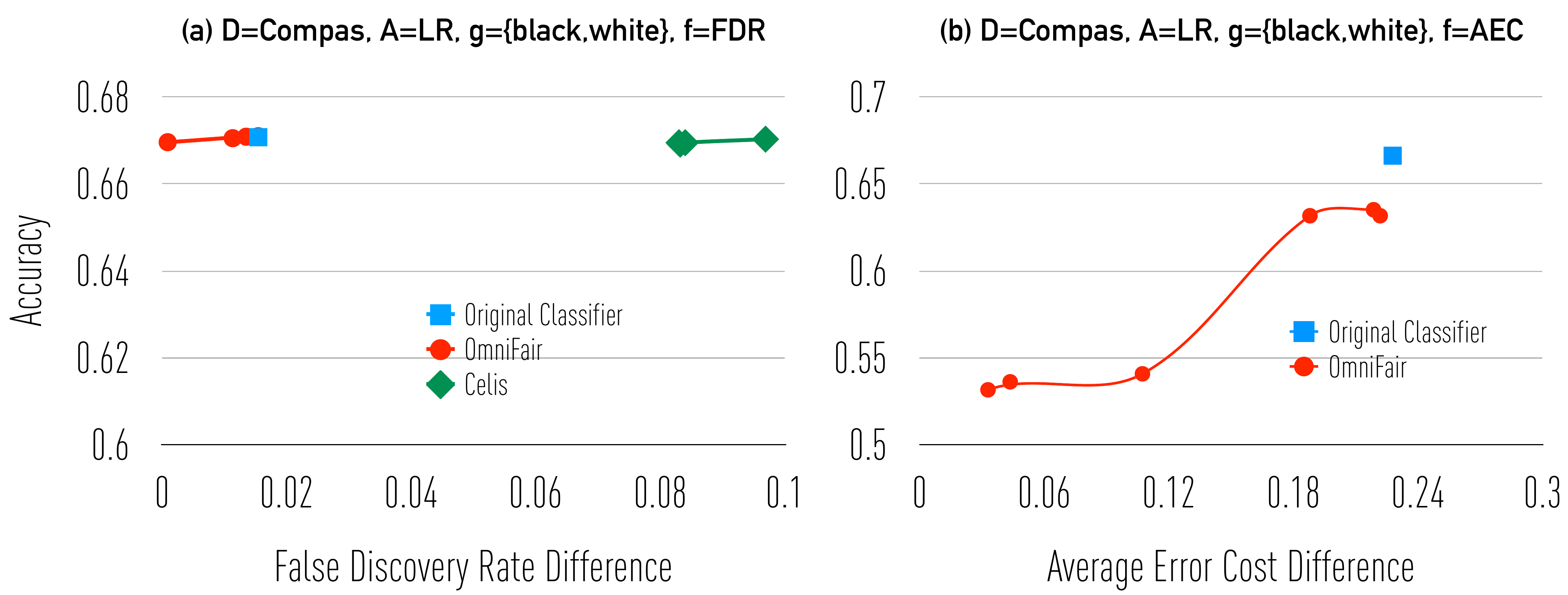}
\caption{\small FDR and AEC on Compas} 
\label{fig:single_fairness_fdr_compas}
\end{figure}

\begin{figure}[h]
\centering
\includegraphics[width=0.6\linewidth]{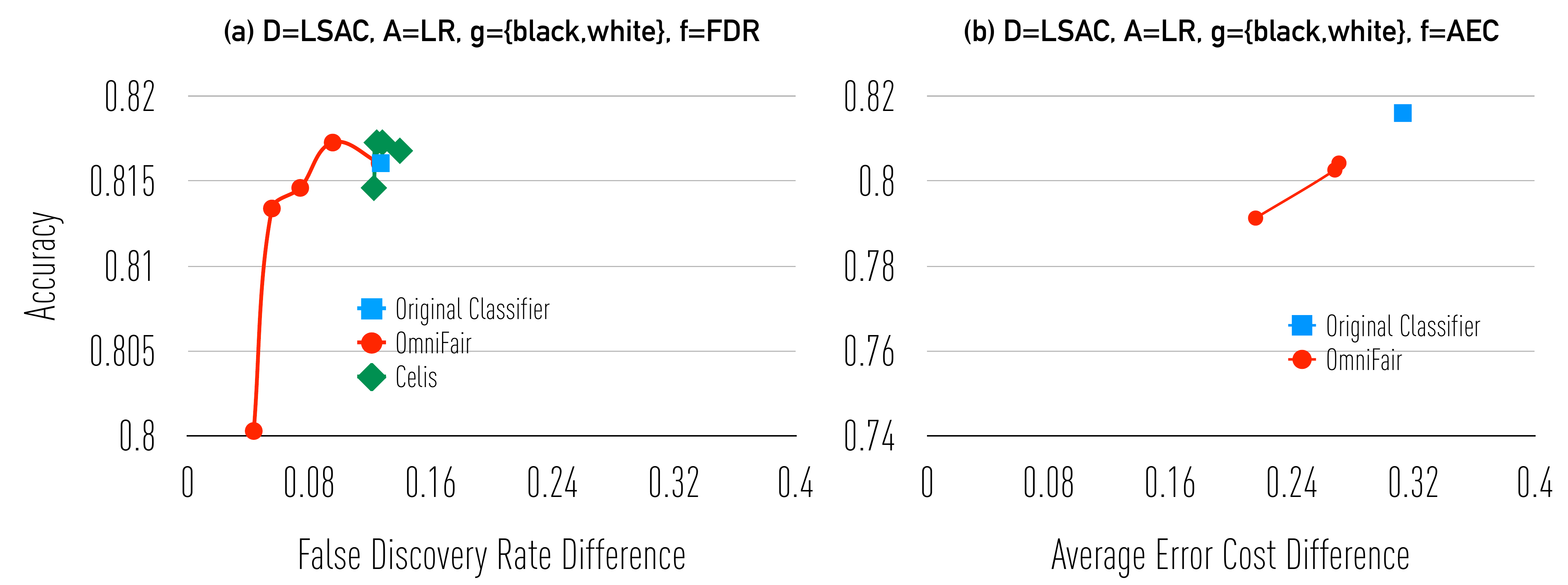}
\caption{\small FDR and AEC on LSAC} 
\label{fig:single_fairness_fdr_lsac}
\end{figure}

\stitle{Accuracy-Fairness Trade-off Comparison by Varying $\varepsilon$.} Similar to SP, we do the analysis of the trade-off between accuracy and fairness by varying $\varepsilon$. Only Celis is able to support False Discovery Rate and \systemx is the only algorithm that supports customized metric. Figure~\ref{fig:single_fairness_fdr_compas} and~\ref{fig:single_fairness_fdr_lsac} show the result for COMPAS dataset and LSAC dataset, respectively. Our algorithm is able to reduce the False Discovery Rate difference
and a customized metric error cost (AEC) while having little accuracy drop.

\end{document}